\documentclass[amsmath,amssymb,aps,showkeys,longbibliography]{revtex4-2}

\usepackage{graphicx}
\usepackage{dcolumn}
\usepackage{bm}
\usepackage[utf8]{inputenc}
\usepackage{textgreek}
\usepackage{palatino}
\usepackage{mathpazo}
\usepackage{textcomp}
\usepackage{subcaption}
\usepackage{ragged2e}
\usepackage{xcolor}
\usepackage{hyperref}
\hypersetup{           
	colorlinks=true,                
	breaklinks=true,                
	urlcolor= blue,                
	linkcolor= magenta,                
	bookmarksopen=false,
	filecolor=black,
	citecolor=red,
	linkbordercolor=blue}
\usepackage{orcidlink}

\begin{document}

\preprint{APS/123-QED}

\title{Gravitational Baryogenesis Constraints on Nojiri–Odintsov $f(R)$ Gravity}

\author{Kalyan Malakar\orcidlink{0009-0002-5134-1553}}
\affiliation{Department of Physics, Dibrugarh University, Dibrugarh, 786004, Assam, India \\ Department of Physics, Silapathar College, Dhemaji, 787059, Assam, India}
 \email{kalyanmalakar349@gmail.com}

\author{Mrinnoy M Gohain\orcidlink{0000-0002-1097-2124}}
\affiliation{
 Department of Physics, Dibrugarh University, Dibrugarh, 786004, Assam, India
}
 \email{mrinmoygohain19@gmail.com}

\author{Kalyan Bhuyan\orcidlink{0000-0002-8896-7691}}
\affiliation{%
Department of Physics, Dibrugarh University, Dibrugarh, 786004, Assam, India \\ Theoretical Physics Division, Centre for Atmospheric Studies, Dibrugarh University, Dibrugarh, 786004, Assam, India}
 \email{kalyanbhuyan@dibru.ac.in}


\begin{abstract}
This paper focuses on exploring the imbalance between matter and antimatter via the gravitational baryogenesis mechanism within the framework of Nojiri-Odintsov  $f(R)$ gravity models in a spatially flat FLRW universe, where $R$ is the Ricci curvature scalar. This mechanism is based on \(\mathcal{CP}\)-violation generated via introducing an interaction between baryonic matter current (\(J^\mu\)) and the derivative of curvature scalar (\(\partial_\mu R\)), which finally results in the baryon asymmetry. We examine three distinct $f(R)$ models: (i) \(f(R)=\frac{-\alpha}{R^m}+\frac{R}{2k^2}+\beta R^2\), (ii) \(f(R)=\alpha R^{\gamma} +\beta R^{\delta}\) and (iii) \(f(R)=\frac{\alpha R^{2m}-\beta R^m}{1+\gamma R^m} \). We demonstrate that even during the matter-dominated epoch of the universe, the $f(R)$ gravity models under consideration can yield a non-vanishing matter-antimatter asymmetry. We constrain the model parameters of each model within the scenario of gravitational baryogenesis to produce the observed value of the baryon-to-entropy ratio, providing a continuous set of acceptable physical values for those parameters. Our results show highly consistent agreement to the baryon-to-entropy ratio with the observational bound from the cosmological data.
\end{abstract}

\maketitle


\section{Introduction}
The asymmetry due to excess of matter over antimatter in the observable cosmos, a conundrum called baryogenesis, has prompted wide-ranging theoretical investigations across many fields of physics, such as particle physics and cosmology. Fundamental physics presumes equal production of particles and antiparticles, resulting in a net zero baryon number \cite{weinberg1995quantum}. Precise cosmological observation data from the Cosmic Microwave Background (CMB) in combination with cosmos's large scale structure (LSS) \cite{bennett2003microwave} and Big Bang Nucleosynthesis (BBN) \cite{burles2001big},  indicate that the observable universe is overwhelmingly dominated by matter, with antimatter confined to extremely rare and highly localized regions \cite{steigman1976observational,dolgov2001matter,cohen1998matter}. The absence of significantly massive antimatter regions, along with the big bang scenario, suggests that the universe evolved from homogeneous baryonic components, pointing towards the necessity for a process such as baryogenesis to account for the observed asymmetry \cite{dine2003origin}. Among the various proposed mechanisms formulated to account for imbalance, gravitational baryogenesis (GB) is characterized to have a dynamical coupling between the baryonic matter current \((J^{\mu})\) and the derivative of Ricci scalar (R) in the evolving spacetime that results in \(\mathcal{CP}\)-violation leading to baryon production \cite{davoudiasl2004gravitational}.

There are three necessary fundamental requirements to be satisfied for small surplus of matter content over antimatter that generated in the early universe as suggested by Sakharov in 1967 as follows \cite{sakharov1998violation}: \\
i.	Baryon number violating interactions in the early universe: At a certain point in the early universe, interactions that violated baryon number were in equilibrium, which led to the generation of asymmetry.\\
ii.	The fundamental laws are violated by Charge Conjugation (\(\mathcal{C}\)) and Charge-Parity (\(\mathcal{CP}\)): This condition is important because net baryon number generation is prevented if \(\mathcal{C}\)- and \(\mathcal{CP}\)- are conserved.\\
iii. Thermal equilibrium violation: Every process and its reverse occur at the same rate when there is thermal equilibrium. Therefore, at some points in the early universe, it must have been out of thermal equilibrium. Thus, an arrow of time exists. 

\cite{davoudiasl2004gravitational} provides the interaction term (coupling between \(J^{\mu}\) and \(\partial_{\mu}R\)) that results in \(\mathcal{CP}\)-violation:

\begin{equation}
\frac{1}{M_*^2}
\int d^4x \,\sqrt{-g}\;J^{\mu}\ (\partial_{\mu}R).
\label{eq:coupling_equation}
\end{equation}
Here, $g$ is the determinant of the metric \(g_{\mu\nu}\), \(M_*\) represents the cut-off scale of the effective field theory that restricts the boundary of the energy regions over which the field will remain effective. 

GB is one of the theoretical techniques developed to explain the observed matter excess in the cosmos. Apart from this mechanism, other procedures have been formulated to explain the matter–antimatter imbalance. Some prominent examples of such competing theories are spontaneous baryogenesis \cite{brandenberger2003spontaneous,takahashi2004spontaneous}, electroweak baryogenesis \cite{trodden1999electroweak}, Affleck–Dine baryogenesis \cite{dine2003origin,stewart1996affleck}, baryogenesis in GUTs (grand unified theories) \cite{kolb1996grand}, and models involving the production of baryons through black hole evaporation \cite{ambrosone2022towards}.

Extended theories of gravity, developed by modifying the Einstein field equations, have offered multifaceted opportunities to explore various aspects of the universe, particularly in the context of contemporary cosmology and so as GB. In 2004, Davoudiasl et al. \cite{davoudiasl2004gravitational} introduced the theoretical mechanism of GB that generated the observed asymmetry via dynamical \(\mathcal{CP}\)-violation. Following that, Lambiase and Scarpetta (2006) \cite{lambiase2006baryogenesis} demonstrated that \(f(R)\) gravity provides a natural arena to generate baryon asymmetry through gravitational baryogenesis. 
In the endeavour to account for the observed Baryon number-to-Entropy Ratio (BnER), \(\frac{\eta}{s}\), a broad range of modified theories of gravity have been deeply investigated by researchers. These alternative theories attempt to go beyond the standard GR framework by invoking curvature-matter couplings, torsion effects, non-metricity contributions, and quantum corrections. The following modified gravity theories are considered in the literature:   \(f(R, T)\) models \cite{sahoo2020gravitational,nozari2018baryogenesis,baffou2019f}, where the action depends on \(R\), the curvature scalar  and \(T\), the trace of the energy-momentum tensor; \(f(R)\) gravity theories \cite{lambiase2006baryogenesis,ramos2017baryogenesis,aghamohammadi2018anisotropy,agrawal2021gravitational}, which generalize Einstein's gravity by some arbitrary functions of \(R\); and the Gauss-Bonnet gravity \cite{odintsov2016gauss}, which includes topological invariants to capture higher-dimensional and string-inspired effects. Other important formulations are \(f(T)\) gravity \cite{oikonomou2016f}, using the framework of the torsion scalar \(T\); and \(f(Q, T)\) gravity \cite{snehasish2020baryogenesis}, where Q stands for non-metricity. Then, hybrid models such as \(f(T, B)\) and \(f(T, T_{G})\) gravity \cite{azhar2020generalized} that involve torsion-based scalars and boundary terms, as well as \(f(R, G)\) and \(f(G, T)\) theories \cite{azhar2021impact}, which include \(G\), the Gauss-Bonnet term. Other formulations are f(P) gravity \cite{bhattacharjee2021baryogenesis}, Modified Horava-Lifshitz gravity \cite{jawad2023viability}, \(f(R, L_{m})\) gravity theory \cite{jaybhaye2023baryogenesis} with \(L_{m}\) being the matter Lagrangian, and quantum fluctuation-corrected gravitational models \cite{yang2024baryogenesis}. Other theories include Extended symmetric teleparallel gravity \cite{narawade2024baryon}, \(f(Q)\) gravity \cite{mishra2024probing}, \(f(Q, C)\) gravity \cite{usman2024compatibility}, where \(C\) is the matter-curvature coupling term, Ricci inverse gravity \cite{jawad2022analysis}, \(f (T, \mathcal{T})\) gravity \cite{mishra2023constraining}, \(f(T, \Theta)\) gravity \cite{rani2022f}, \(\Theta\) being the contraction of the energy-momentum tensor, Einstein-Ather theory \cite{sultan2025observational}, Extended \(f(Q, L_{m})\) gravity \cite{samaddar2025novel}, modified \(f(Q, B)\) gravity \cite{alruwaili2025probing}, Energy-momentum squared gravity \cite{pereira2024gravitational} and Scalar-tensor theories \cite{pereira2024scalar}. All together, these theories offer different mechanisms for baryogenesis in the context of early universe cosmology. They cover both minimal coupling regimes, where the matter sector couples weakly with the gravitational field, and non-minimal coupling frameworks, which involve strong matter-curvature or matter-geometry couplings.

$f(R)$ gravity addresses many crucial aspects of late-time accelerated expansion effectively by imitating the behavior of the \(\Lambda\)CDM model. The $f(R)$ gravity models proposed by Nojiri and Odintsov successfully reproduce the major phases of cosmic evolution, including the radiation and matter-dominated eras, the shift from decelerated to accelerated expansion, and the present-day cosmic acceleration, all in agreement with the three-year WMAP observations \cite{nojiri2003modified,nojiri2006modified,nojiri2008modified}. Based on that, we are motivated to explore whether the $f(R)$ gravity models that offer a coherent framework that naturally links early-universe inflation to the current accelerated expansion can also account for the phenomenon of GB in the universe within an isotropic background. Moreover, analysis within the Newtonian limit shows that these models are compatible with local gravitational tests, such as adherence to Earth-like gravitational solutions, consistency with Newton’s law and ensuring an extra scalar field that is significantly massive, thus positioning them as viable alternatives to GR \cite{nojiri2006modified,nojiri2008modified}.

The paper is arranged as follows: A detailed discussion on \(f(R)\) gravity framework and derivation of the field equation is included in Section \ref{sec2}. Section \ref{sec3} discusses the formulation of GB within the domain of modified \(f(R)\) theories under consideration, along with the corresponding analytical computations. The graphical representations, as well as tabular forms enumerating the theoretical results, have also been furnished herein. Section \ref{sec3} also focuses upon the analysis and interpretation of essential results drawn within the proposed theoretical framework. And finally, in Section \ref{sec4} we summarize the results of our study.

\section{f(R) gravity and Field Equations}
\label{sec2}
\subsection{Description of f(R) gravity and FLRW metric}
Over the past two decades, numerous observational evidences have established that the cosmos is going through an accelerated expansion, particularly in its late-time evolution \cite{riess1998observational,spergel2003first}. This phenomenon has underlined the limitation of GR. This challenge has driven the evolution of extended theories gravity, which either extend the framework of GR or propose alternative formulations. In response, many gravity theories have emerged that have a viable mechanism to account for cosmic acceleration without requiring exotic dark energy components. Thus, the modification provides a robust framework for analysing late-time cosmological dynamics \cite{carroll2004cosmic}. The modified action for \(f(R)\) gravity that incorporates both the geometric contribution from the modified curvature term and the matter content of the universe reads as \cite{lambiase2006baryogenesis}: 

\begin{equation}
S_{fR}=
\frac{1}{2k^2}\int \sqrt{-g}d^4x \,f(R) + \int\sqrt{-g} d^4x \,L_{m}(\psi,g_{\mu\nu})
\label{eq:action_equation}
\end{equation}

here, \(\psi\) and \(L_{m}(\psi,g_{\mu\nu})\) are the Matter field and Lagrangian of the matter field, respectively. \(k^2\) is a positive constant, \(k^2=8\pi G\). Natural unit system is taken into consideration throughout the paper: \(c=\hbar=8\pi G=1\).

The field equations can be derived by taking variation of eqn. \eqref{eq:action_equation} w.r.t. metric:

\begin{equation}
R_{\mu\nu}-\frac{1}{2}g_{\mu\nu}f^{\prime}(R)-(\nabla_{\mu}\nabla_{\nu}-g_{\mu\nu}\Box)f^{\prime}(R)=T_{\mu\nu}
\label{eq:general_field_equation}
\end{equation}

Prime over \(f(R)\) represents differentiation of \(f(R)\) w.r.t. \(R\).

The matter energy-momentum tensor (\(T_{\mu\nu}\)) can be expressed as follows:  

\begin{equation}
T_{\mu\nu}=-2\frac{1}{\sqrt{g}}\frac{\delta(\sqrt{-g}L_{m})}{\delta(g^{\mu\nu})}
\label{eq:energy-momentum_tensor}
\end{equation}

The trace of the eqn. \eqref{eq:general_field_equation} can be obtained by contracting the field equation and the resulting form is: 

\begin{equation}
f^{\prime}(R).R+3\Box f^{\prime}(R)-2f(R)=T
\label{eq:trace_field_equation}
\end{equation}

In this formalism, the energy-momentum conservation condition is given by the following condition: 

\begin{equation}
\nabla_{\mu}(T^{\mu\nu})=0
\label{eq:conservation_energy_momentum}
\end{equation}

Under the assumptions of large-scale homogeneity and isotropy, the universe's geometry is demonstrated by the Friedmann–Lemaître–Robertson–Walker (FLRW) metric that acts as the paradigm for present-time cosmological models \cite{rasanen2015new},

\begin{equation}
ds^2=dt^2-a^2(t)(\frac{dr^2}{1-Kr^2}+r^2d\theta+r^2Sin^2\theta d\phi^2)
\label{eq:FLRW_metric}
\end{equation}

here, (\(r,\theta,\phi\)) is the coordinates of the spherical coordinate system at cosmic time t. \(K\) represents the curvature of the space. \(a(t)\) being the scale factor of the accelerating universe. Spatial Curvature (\(K\)) in the metric eqn. \eqref{eq:FLRW_metric} may not only be positive, but also zero or negative, leading to the possibilities of a closed, flat and open universe.

For this analysis, we adopt an isotropic and spatially flat universe \((K=0)\), leading to a modified form of metric eqn. \eqref{eq:FLRW_metric} as follows:

\begin{equation}
ds^2=dt^2-a^2(t)(dx^2+dy^2+dz^2)
\label{eq:flat_FLRW_metric}
\end{equation}

The preference for a spatially flat universe in eqn. \eqref{eq:flat_FLRW_metric} arises from the observations of baryon acoustic oscillations (BAO), CMB measurements and large-scale structure (LSS) surveys \cite{l2017model,jimenez2019measuring, foidl2024lambda}, which helps us comprehend the expansion history of the cosmos and its current accelerated phase. In this paper, we will constrain our study within a homogeneous and isotropic spatially flat universe (\(K=0\)).

\subsection{Field equations}
Under the idealization of matter as a perfect fluid, the energy-momentum tensor, which encapsulates the distribution and flow of energy and momentum in spacetime, is expressed as:

\begin{equation}
T_{\mu\nu}=(\rho+p)u_{\mu}u_{\nu}-pg_{\mu\nu}
\label{eq:T_as_perfect_fluid}
\end{equation}

\(u^{\mu}\) is the perfect fluid four-velocity vector that satisfies the criterion, \(u^{\mu}u_{\mu}=1\). \(\rho\) and \(p\) stands for the perfect fluid element's energy density and pressure, respectively. 

Friedmann Field Equations (FFE's) can be reduced from the field eqn. \eqref{eq:general_field_equation} for a flat \(\&\) isotropic universe as follows: 

\begin{eqnarray}
&&3H\dot{R}f^{\prime\prime}(R)-3(\dot{H}+H^2)f^{\prime}(R)-\frac{1}{2}f = \rho 
\label{eq:friedmann_equation1} \\&&
-\dot{R}f^{\prime\prime\prime}(R)-(\ddot{R}+2H\dot{R})f^{\prime\prime}(R)+(\dot{H}+3H^2)f^{\prime}(R)+\frac{1}{2}f(R)=p
\label{eq:friedmann_equation2}
\end{eqnarray}

The energy conservation condition can be derived from the Bianchi identities and is given by:

\begin{equation}
3(p+\rho)H+\dot{\rho}=0
\label{eq:energy_conservation}
\end{equation}

For a flat FLRW metric, R analytically yields the following result:

\begin{equation}
R=-6(2H^2+\dot{H})
\label{eq:ricci_scalar}
\end{equation}

In order to evaluate solutions to eqns. \eqref{eq:friedmann_equation1} and \eqref{eq:friedmann_equation2}, it is essential to consider a specific $f(R)$ form rather than its generalized representation. We propose exploring the modified $f(R)$ models proposed by S. Nojiri \& S. D. Odintsov \cite{nojiri2006modified,nojiri2008modified} designed to unify the early-time inflation epoch, radiation \& matter dominated era, and late-time accelerated expansion era \cite{nojiri2008dark}.  We will test whether these models can generate the observed baryon asymmetry via a \(\mathcal{CP}\)-violating interaction between $R$ and the baryon current density \cite{davoudiasl2004gravitational}. By deriving the time evolution of $R$ in each model considered and calculating the induced BnER, we will identify the model parameter regions where successful GB is achieved \cite{nojiri2006modified,samanta2020physical}.

Modified $f(R)$ gravity models where cosmic unification is present have yet to demonstrate their ability to generate the observed baryon asymmetry ratio via GB. Thus, the motivation of our work is to study the process of GB in the specified $f(R)$ models that unify the early-time inflation epoch, the matter and radiation-dominated era, and the late-time universe acceleration epoch.

\subsection{Gravitational Baryogenesis}

According to current astrophysical understandings, the Big Bang is expected to have produced matter and antimatter in the same amounts, resulting in a net zero baryon number. However, cosmological observations \cite{burles2001big,bennett2003microwave}, along with the absence of significant matter-antimatter dissolution radiation \cite{cohen1998matter}, indicate a clear prevalence of matter over anti-matter in the universe. To quantify this imbalance, a dimensionless quantity is defined as follows: 

\begin{equation}
\frac{\eta}{s}=\frac{n_{\beta}-n_{\bar\beta}}{s}
\label{eq:baryon-number-to-entropy}
\end{equation}

\(n_{\beta}\) and \(n_{\bar{\beta}}\) are the baryon and anti-baryon number densities, respectively, and s represents the entropy of the universe.

It can be confirmed from the observations of CMB \cite{burles2001big} and BBN \cite{bennett2003microwave} that the magnitude of the asymmetry is; \(\frac{\eta}{s}\cong 9.42 \times 10^{-11}\).

The coupling eqn. \eqref{eq:coupling_equation} enforces \(\mathcal{CP}\)-violation that dynamically generates a non-zero asymmetry value by preserving a non-vanishing \(\dot{R}\). Once the temperature of the cosmos falls below a critical point called the decoupling temperature (\(T_D\)), the \(\mathcal{CP}\)-violation disappears, leading to a net matter abundance over antimatter. Thus, the interaction that produces the imbalance occurs at \(T=T_D\). Thereafter, the asymmetry sustained in the cosmos is nearly equal to \cite{davoudiasl2004gravitational}:

\begin{equation}
\frac{\eta}{s}=-\frac{15}{4\pi^2}\frac{g_b}{g_*}\frac{\dot{R}}{M_*^2T_D}
\label{eq:general_baryontoentropyratio}
\end{equation}

here, \(g_b\) accounts for the baryonic content's total intrinsic degrees of freedom (d.o.f). \(g_*\) is the effective number of d.o.f of massless particles.

The evolution of the universe from one equilibrium energy state to another is fundamentally governed by the relationship between energy and temperature. For the universe in thermal equilibrium, the energy density \(\rho\) is directly related to the temperature ($T$) as \cite{piattella2018lecture}:

\begin{equation}
\rho=\frac{\pi^2}{30}g_*T^4
\label{eq:energy_density}
\end{equation}

As the universe expands, it cools, leading to a decrease in temperature and energy density. This cooling drives the universe through a series of phase transitions, each corresponding to a new equilibrium state characterized by different symmetries and particle interactions. These transitions include significant events such as \(\mathcal{CP}\)-violation that leads to the generation of baryon asymmetry, which have profound implications for the formation of fundamental particles and the LSS of the cosmos \cite{sugamoto1995baryon,balaji2005dynamical,huber2023baryogenesis}.
To further solve the aforementioned equations, we adopt the scale factor of the form: 
\begin{equation}
a(t)=a_0t^n
\label{eq:scalefactor}
\end{equation}
where, $n$ is a real number constrained to positive values \((n > 0)\). 

The choice of a general power law form scale factor, \(a(t)=a_0 t^n\),  provides an analytically traceable wide-range framework to study GB. The specific values of \(n\) correspond to well-known epochs like radiation-dominated \(n=\frac{1}{2}\) or matter-dominated \(n=\frac{2}{3}\), etc. Moreover, the general power-law form allows one to explore baryon asymmetry generation across a broader class of cosmological backgrounds. This is particularly useful in the regime of $f(R)$ theories, where such power-law expansions naturally emerge as exact solutions. Thus, it offers a unified approach that accommodates standard epochs as special cases while preserving generality in the analysis.

Using the scale factor, \eqref{eq:scalefactor}, we can calculate the Hubble parameter and Ricci scalar as follows: 
\begin{equation}
 H=\frac{n}{t}, \hspace{4cm} R=\frac{6 n (2 n-1)}{t^2}
\end{equation}

\section{Models of f(R) gravity and Gravitational Baryogenesis}
\label{sec3}
We propose exploring the GB mechanisms in certain modified \(f(R)\) gravity theories, specifically those by S. Nojiri and S. D. Odintsov \cite{nojiri2006modified,nojiri2008modified} in a flat isotropic universe. We will investigate whether the Nojiri–Odintsov \(f(R)\) models can naturally explain the observed asymmetry ratio of baryon number-to-entropy via GB and thereby advance the current understanding of both modified gravity and the cosmology of the early universe. The procedure involves the introduction of a coupling term between the derivative of $R$ and the baryon current density \(J^{\mu}\) \cite{davoudiasl2004gravitational}. The analysis further investigates how the ratio rely on the functional form of $f(R)$ under consideration.  We constrain the model parameters associated with the models within a physically acceptable range by aligning the theoretical predictions with the observed cosmological data. This study aims to assess the feasibility of these modified gravity scenarios as sources of GB and to shed light on how curvature corrections to GR, unifying early inflation and late-time acceleration, can influence particle-physics phenomena in the early universe. In subsequent analysis, we first calculate the decoupling time (t\textsubscript{D}) as a function of the decoupling temperature (T\textsubscript{D}) and then evaluate the BnER for all three models under investigation.

\subsection{Model I}
The first modified gravity model we have considered here to study GB is \cite{nojiri2006modified}:
\begin{eqnarray}
f(R)=\frac{-\alpha}{R^m}+\frac{R}{2k^2}+\beta R^2
\end{eqnarray}
Here, \(\alpha\) and \(\beta\) are free model parameters, $m$ is a positive integer. \\

This model has a diverse functional form that incorporates early-time inflation, \cite{cuzinatto2024inflationary,tomita2016note}, the radiation- and matter-dominated era \cite{arik2010radiation} as well as late-time acceleration expansion \cite{rusyda2024constraining,arora2023late} within its framework \cite{nojiri2003modified}. These combined features form a self-consistent $f(R)$ model that provides an elegant and effective theoretical setting for our study of GB.

To solve eqns. \eqref{eq:friedmann_equation1} and \eqref{eq:friedmann_equation2} with analytical methods for this particular model, we will have to give some specified values to $m$. In our work, we take the cases \(m=1\) and \(m=2\).

\subsubsection{Case I}
For Case I (\(m=1\)), the $f(R)$ functional form modifies as follows: 
\begin{eqnarray}
f(R)=\frac{-\alpha}{R}+\frac{R}{2k^2}+\beta R^2
\end{eqnarray}
Using \(f^{\prime}\) and \(f^{\prime\prime}\) in the first motion eqn. \eqref{eq:friedmann_equation1}, the expression of the energy density can be approximated as follows: 
\begin{equation}
 -\frac{2 \beta  n^3 (2 n-1)^3 (14 n+11)+n^2 (8 n-7) (1-2 n)^2 t^2+2 \alpha  (4 n-29) t^6}{4 (1-2 n)^2 n t^4}
 \label{eq:energy_density_case1_model1}
\end{equation}
The expression of the energy density in eqn. \eqref{eq:energy_density_case1_model1} upon equating with the energy density in eqn. \eqref{eq:energy_density}, the decoupling time \((t_D)\) can be expressed as a function of the  decoupling temperature \((T_D)\):
\begin{equation}
t_D = \sqrt{\frac{\sqrt[3]{2} A^2}{3 F \sqrt[3]{J}}-\frac{A}{3 F}-\frac{\sqrt[3]{2} B}{\sqrt[3]{J}}+\frac{\sqrt[3]{J}}{3 \sqrt[3]{2} F}}
\label{eq:tD_for_model1_case1}
\end{equation}
where
\begin{align*}
J &= \left(G + \sqrt{G^2 + 4 L^3}\right),
\hspace{0.5cm} G = \left(-2 A^3+9 A B F-27 C F^2\right), \hspace{0.5cm} L = \left(3 B F-A^2\right), \\
A &=  \left(\frac{8}{15} \pi ^2 g_{*} n^3 T_D^4-\frac{8}{15} \pi ^2 g_{*} n^2 T_D^4+\frac{2}{15} \pi ^2g_{*} n T_D^4\right),
\hspace{0.1cm} B = \left(32 n^5-60 n^4+36 n^3-7 n^2\right), \\
C &= \left(224 \beta  n^7-160 \beta  n^6-96 \beta  n^5+104 \beta  n^4-22 \beta  n^3\right), \hspace{0.1cm} D =\left(8 \alpha  n-58 \alpha\right).
\end{align*}
Inserting eqns. \eqref{eq:ricci_scalar} and \eqref{eq:tD_for_model1_case1} into the eqn. \eqref{eq:general_baryontoentropyratio}, the BnER \((\frac{\eta}{s})\) can be derived mathematically as follows: 

\begin{equation}
\frac{\eta}{s} = \frac{15 g_b}{2 \pi^2 g_{*} M_*^2  T_D}\frac{(2n - 1)n}{S^{3/2}}
\label{eq:ratio_model1_case1}
\end{equation}
where
\begin{align*}
\mathcal{S} &= \left(\frac{\sqrt[3]{2} \mathcal{A}^2}{3 \mathcal{D} \sqrt[3]{\sqrt{4 \mathcal{N}^3+\mathcal{M}^2}+\mathcal{M}}}-\frac{\mathcal{A}}{3 \mathcal{D}}+\frac{\sqrt[3]{\sqrt{4 \mathcal{N}^3+\mathcal{M}^2}+\mathcal{M}}}{3 \sqrt[3]{2} \mathcal{D}}-\frac{\sqrt[3]{2} \mathcal{B}}{\sqrt[3]{\sqrt{4 \mathcal{N}^3+\mathcal{M}^2}+\mathcal{M}}}\right),\\
\mathcal{M} &= \left(-2 \mathcal{A}^3+9 \mathcal{A} \mathcal{D} \mathcal{B}-27 \mathcal{C} \mathcal{D}^2\right), \hspace{0.5cm} \mathcal{A} = \left(\frac{8}{15} \pi ^2 g_* n^3 T_D^4-\frac{8}{15} \pi ^2 g_* n^2 T_D^4+\frac{2}{15} \pi ^2 g_* n T_D^4\right),\\
\mathcal{N} &= \left(3 \mathcal{D} \mathcal{B}-\mathcal{A}^2\right), 
\hspace{2.7cm} \mathcal{B} = \left(-7 n^2 + 36 n^3 - 60 n^4 + 32 n^5\right),\\
\mathcal{C} &= \left(-22 n^3 \beta + 104 n^4 \beta - 96 n^5 \beta - 
  160 n^6 \beta + 224 n^7 \beta\right), \hspace{0.5cm} \mathcal{D} =\left(8 \alpha  n-58 \alpha\right).
\end{align*}

\begin{figure*}[htbp]
\centering
\begin{subfigure}[b]{0.45\textwidth}
  \centering
   \hspace*{-1cm} 
  \includegraphics[width=1.10\linewidth, trim=0cm 0 1cm 1cm, clip]{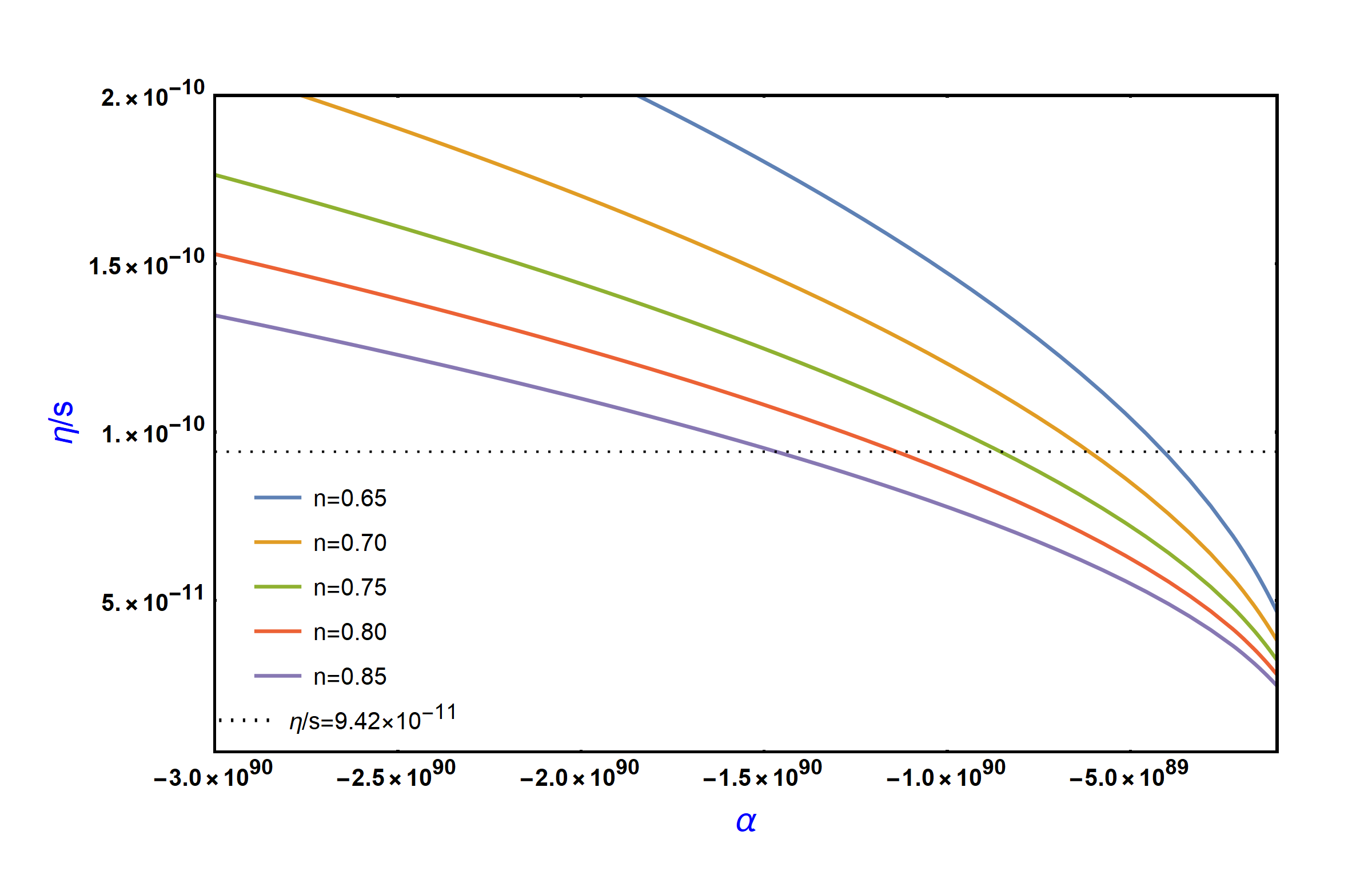}
  \caption{\justifying Plot of \(\frac{\eta}{s}\) versus \(\alpha\) for different values of \(n\) with \(g_b \simeq 1\), \(M_*=1\times10^{12} GeV\), \(T_D=2\times10^{16} GeV\), \(g_*=106\) and \(\beta = -4 \times 10^{25}\).}
  \label{fig:Figure1}
\end{subfigure}
\hfill
\begin{subfigure}[b]{0.45\textwidth}
  \centering
   \hspace*{-0.5cm} 
  \includegraphics[width=1.10\linewidth, trim=0cm 0 1cm 1cm, clip]{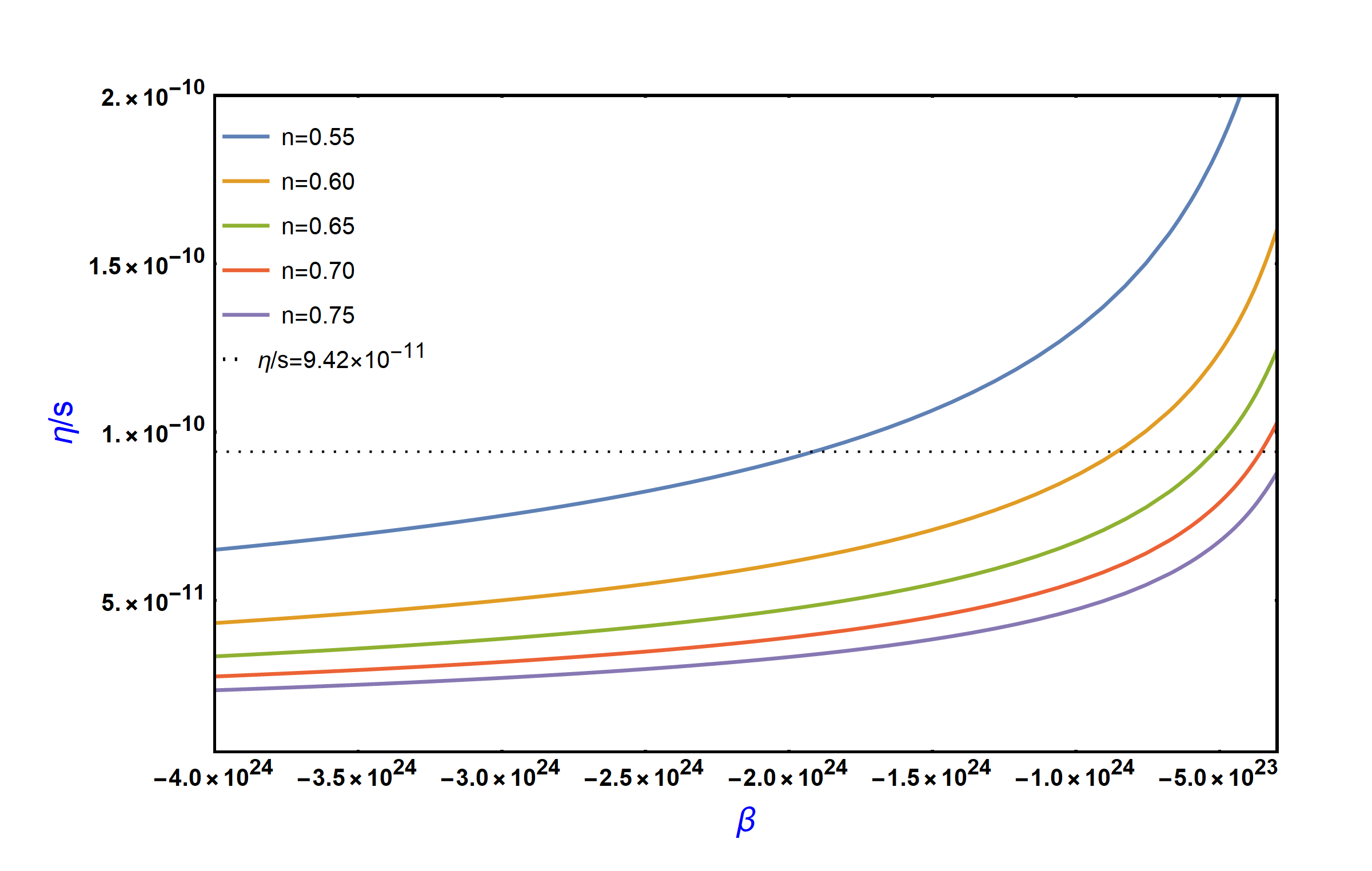}
  \caption{\justifying Plot of \(\frac{\eta}{s}\) versus \(\beta\) for different values of \(n\) with \(g_b \simeq 1\), \(M_*=1\times10^{12} GeV\), \(T_D=2\times10^{16} GeV\), \(g_*=106\) and \(\alpha = -5 \times 10^{87}\).}
  \label{fig:Figure2}
\end{subfigure}

\vspace{0.5cm}
\begin{subfigure}[b]{0.45\textwidth}
  \centering
   \hspace*{-1cm} 
  \includegraphics[width=1.10\linewidth, trim=0cm 0 1cm 1cm, clip]{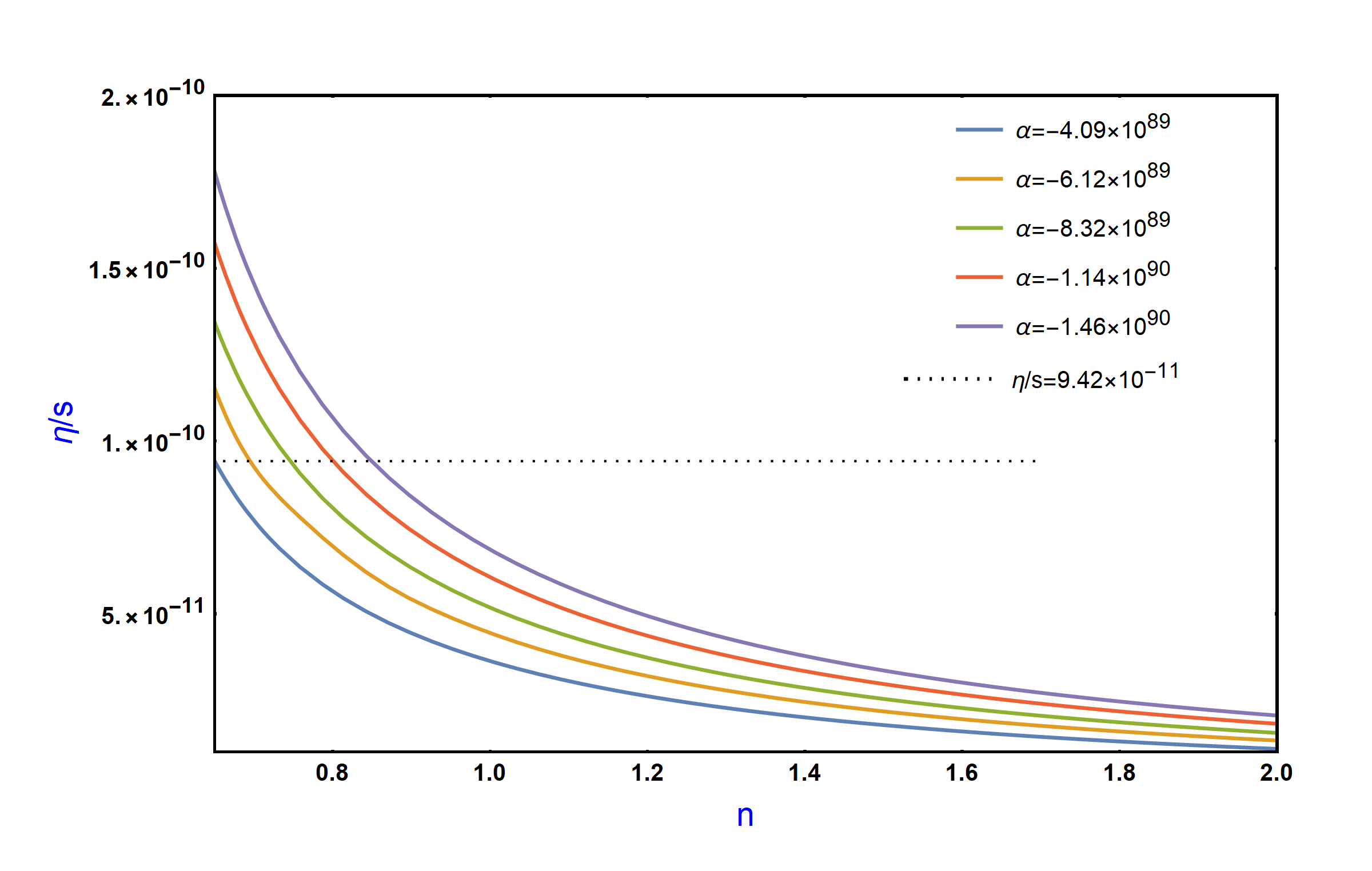}
  \caption{\justifying Plot of asymmetry ratio versus \(n\) for different values of \(\alpha\) with \(g_b \simeq 1\), \(M_*=1\times10^{12} GeV\), \(T_D=2\times10^{16} GeV\), \(g_*=106\) and \(\beta = -4 \times 10^{25}\).}
  \label{fig:Figure3}
\end{subfigure}
\hfill
\begin{subfigure}[b]{0.45\textwidth}
  \centering
   \hspace*{-0.5cm} 
  \includegraphics[width=1.10\linewidth, trim=0cm 0 1cm 1cm, clip]{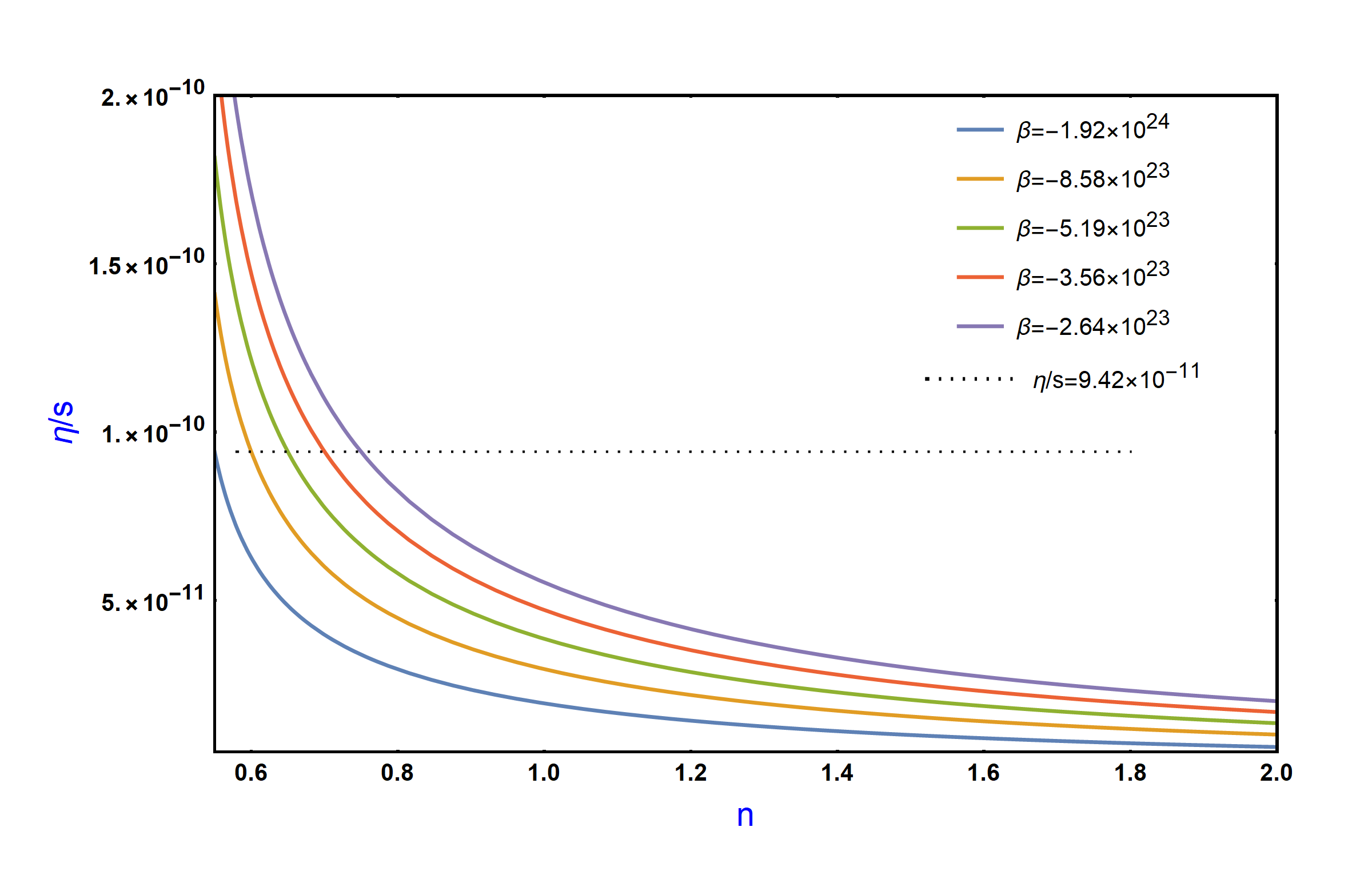}
  \caption{\justifying Plot of asymmetry ratio versus \(n\) for different values of \(\beta\) with \(g_b \simeq 1\), \(M_*=1\times10^{12} GeV\), \(T_D=2\times10^{16} GeV\), \(g_*=106\) and \(\alpha = -5 \times 10^{87}\).}
  \label{fig:Figure4}
\end{subfigure}

\caption{Plot of \(\frac{\eta}{s}\) as a function of \(\alpha\), \(\beta\) and \(n\) for Case I (\(m=1\)) of Model I.}
\label{fig:fig_case1_model1}
\end{figure*}

The BnER resulting from eqn. \eqref{eq:ratio_model1_case1} is non-zero, unlike that in the GR if \(n \neq \frac{1}{2}\). In the case of \(n=\frac{1}{2}\), eqn. \eqref{eq:ratio_model1_case1} becomes singular. Thus, \(n=\frac{1}{2}\) lies outside the range of valid parameters for our model. As pointed in \cite{davoudiasl2004gravitational}, it is natural to have interaction terms like \eqref{eq:coupling_equation} in the low-energy effective field approximation derived from more fundamental quantum field theory (QFT) in curved spacetime, provided the cut-off scale \((M_*)\) is near around the reduced Planck mass \((M_p)\). The model can produce the necessary asymmetry if the temperature \((T_D)\) at which the asymmetry is generated is equal to \(M_I\), the inflation scale. The value of \(M_I\)  is about \(2\times10^{16}\) GeV, which is based on the gravitational waves detected by LIGO \cite{oikonomou2016f,lambiase2006baryogenesis,ramos2017baryogenesis}. The values of the constant parameters of eqn. \eqref{eq:ratio_model1_case1} are \(g_b \simeq 1\), \(g_*=106\), \(M_*=1\times10^{12} GeV\) and \(T_D=2\times10^{16} GeV\) \cite{lambiase2006baryogenesis,sahoo2020gravitational,nozari2018baryogenesis,baffou2019f,ramos2017baryogenesis,aghamohammadi2018anisotropy,agrawal2021gravitational,odintsov2016gauss,oikonomou2016f,snehasish2020baryogenesis,azhar2020generalized,azhar2021impact,bhattacharjee2021baryogenesis,jawad2023viability,jaybhaye2023baryogenesis}. In fig. \ref{fig:Figure1}, the asymmetry ratio \(\frac{\eta}{s}\) is plotted against \(\alpha\) for five different values of \(n\) which are represented by different colors, taking \(\beta\) constant, \(\beta=-4\times10^{25}\). \autoref{tab:tableA} provides a set of values of  \(n\) , \(\alpha\) and \(\frac{\eta}{s}\), for which the theoretically calculated values of the asymmetry ratio are very close to the measured values of the ratio, \(9.42\times10^{-11}\), from cosmological observations. Fig. \ref{fig:Figure2}, \(\frac{\eta}{s}\) is plotted against \(\beta\) for different values of \(n\), making \(\alpha\) fixed, \(\alpha=-5\times10^{87}\). Fig. \ref{fig:Figure3} and fig. \ref{fig:Figure4} illustrates the dependence of the baryon asymmetry ratio with respect to \(n\) for varying \(\alpha\) and \(\beta\) values, respectively. The horizontal dotted line represents the observed value of the ratio. 

\begin{table}[htb]
  \caption{For Case I (\(m = 1\)) of Model I, by fixing \(\beta = -4 \times 10^{25}\), the following table includes a set of values of \(n\) and \(\alpha\) for which the generated matter imbalance ratio have values close to that of the observed asymmetry ratio.}
  \begin{tabular}{c@{\hspace{1cm}}c@{\hspace{1cm}}c}
    \hline \hline
    $n$ & $\alpha$ & $\frac{\eta}{s}$ \\
    \hline
    0.7 & $-6.12 \times 10^{89}$ & $9.417 \times 10^{-11}$ \\
    0.8 & $-1.14 \times 10^{90}$ & $9.428 \times 10^{-11}$ \\
    0.9 & $-1.84 \times 10^{90}$ & $9.412 \times 10^{-11}$ \\
    \hline \hline
  \end{tabular}
  \label{tab:tableA}
\end{table}

\subsubsection{Case II}
For Case II (\(m=2\)), the \(f(R)\) functional form modifies as follows:
\begin{equation}
f(R)=\frac{-\alpha}{R^2}+\frac{R}{2k^2}+\beta R^2
\end{equation}

\begin{figure*}[htbp]
\centering
\begin{subfigure}[b]{0.45\textwidth}
  \centering
   \hspace*{-1cm} 
  \includegraphics[width=1.10\linewidth, trim=0 0 1cm 1cm, clip]{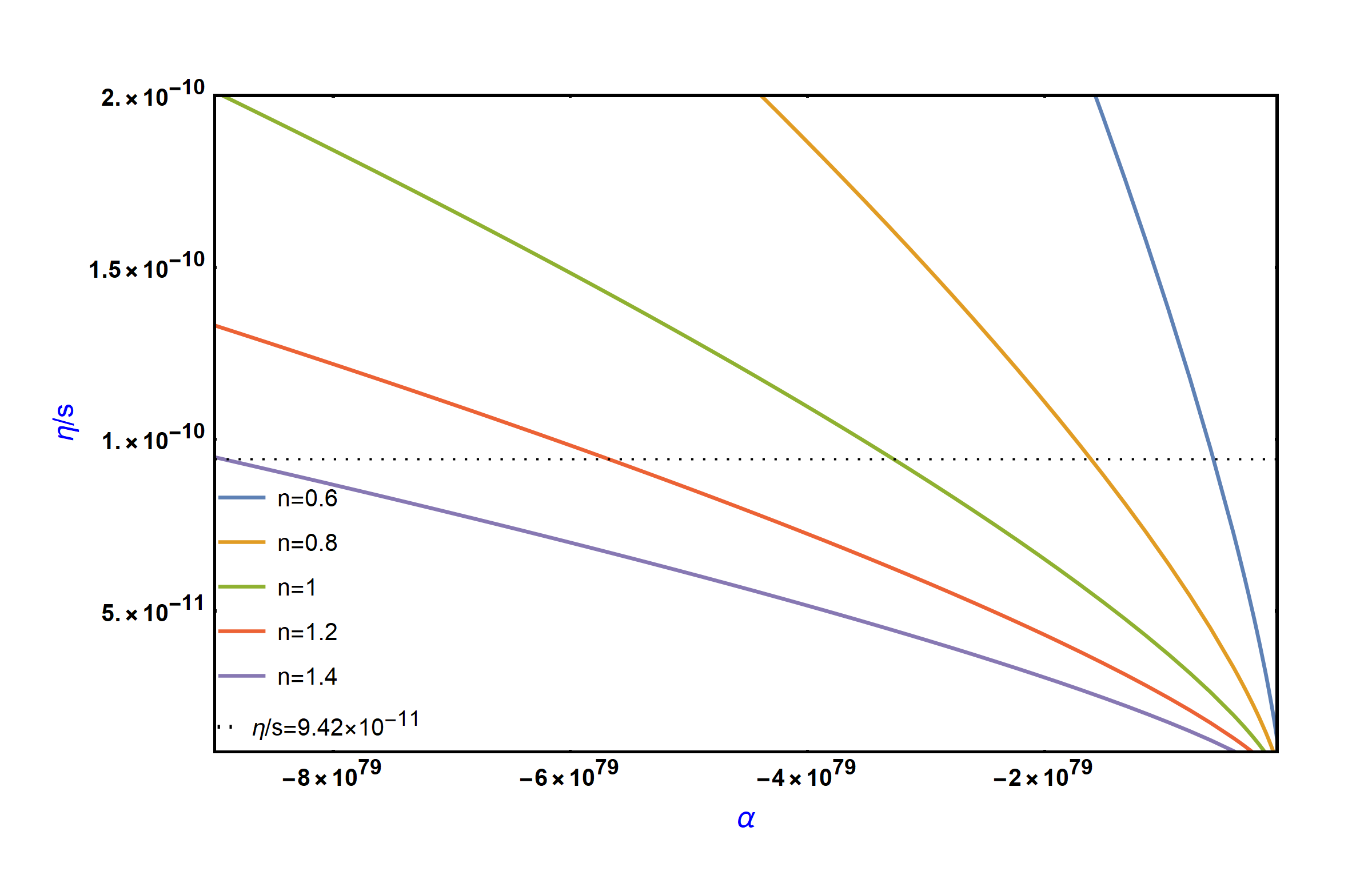}
  \caption{\justifying Plot of \(\frac{\eta}{s}\) versus \(\alpha\) for different values of \(n\) with \(g_b \simeq 1\), \(M_*=1\times10^{12} GeV\), \(T_D=2\times10^{16} GeV\), \(g_*=106\) and \(\beta=1\).}
  \label{fig:Figure6}
\end{subfigure}
\hfill
\begin{subfigure}[b]{0.45\textwidth}
  \centering
   \hspace*{-0.5cm} 
  \includegraphics[width=1.10\linewidth, trim=0 0 1cm 1cm, clip]{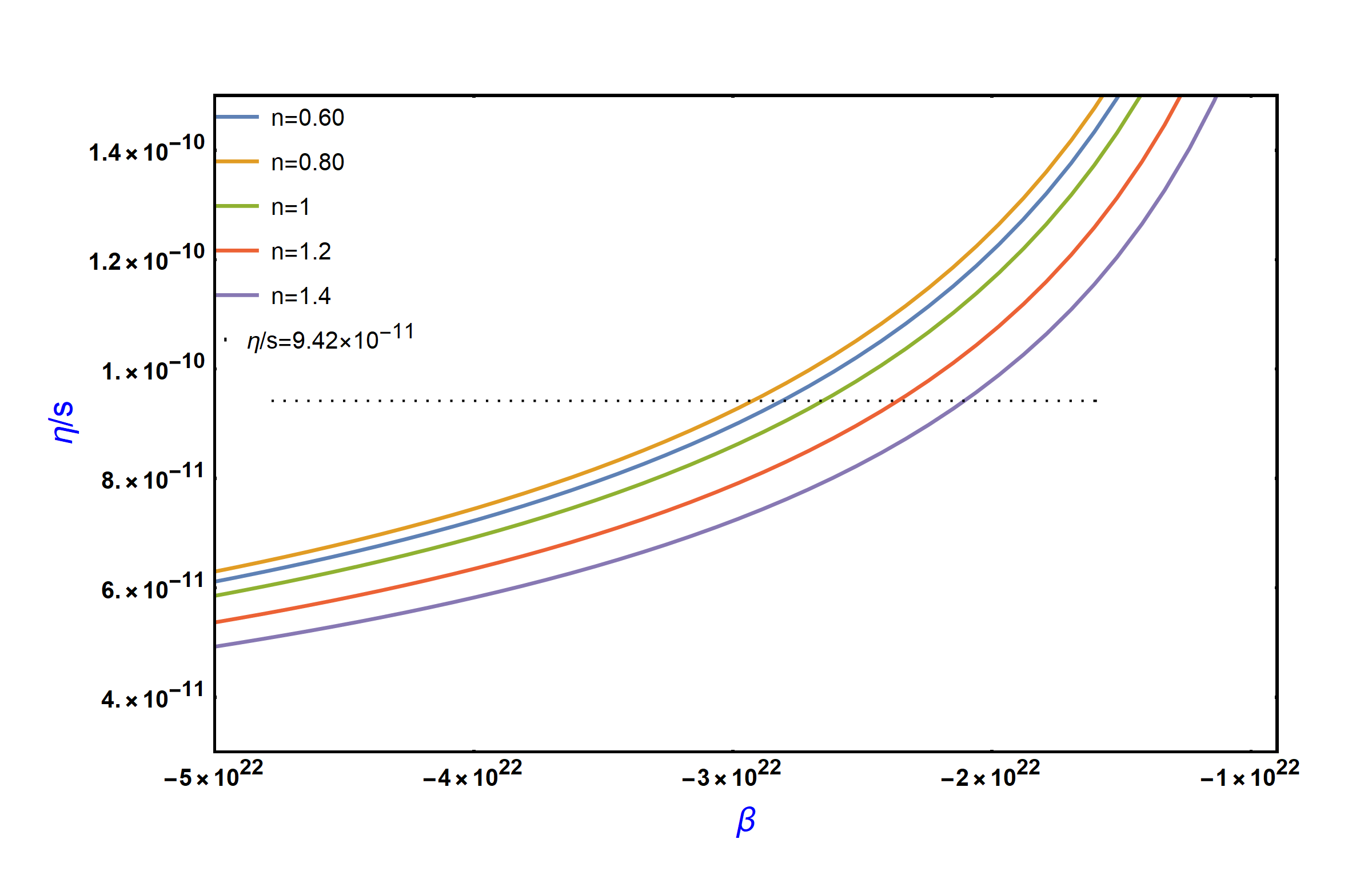}
  \caption{\justifying Plot of \(\frac{\eta}{s}\) versus \(\beta\) for different values of \(n\) with \(g_b \simeq 1\), \(M_*=1\times10^{12} GeV\), \(T_D=2\times10^{16} GeV\), \(g_*=106\) and \(\alpha = -5 \times 10^{96}\).}
  \label{fig:Figure7}
\end{subfigure}

\vspace{0.5cm}

\begin{subfigure}[b]{0.45\textwidth}
  \centering
   \hspace*{-1cm} 
 \includegraphics[width=1.10\linewidth, trim=0 0 1cm 1cm, clip]{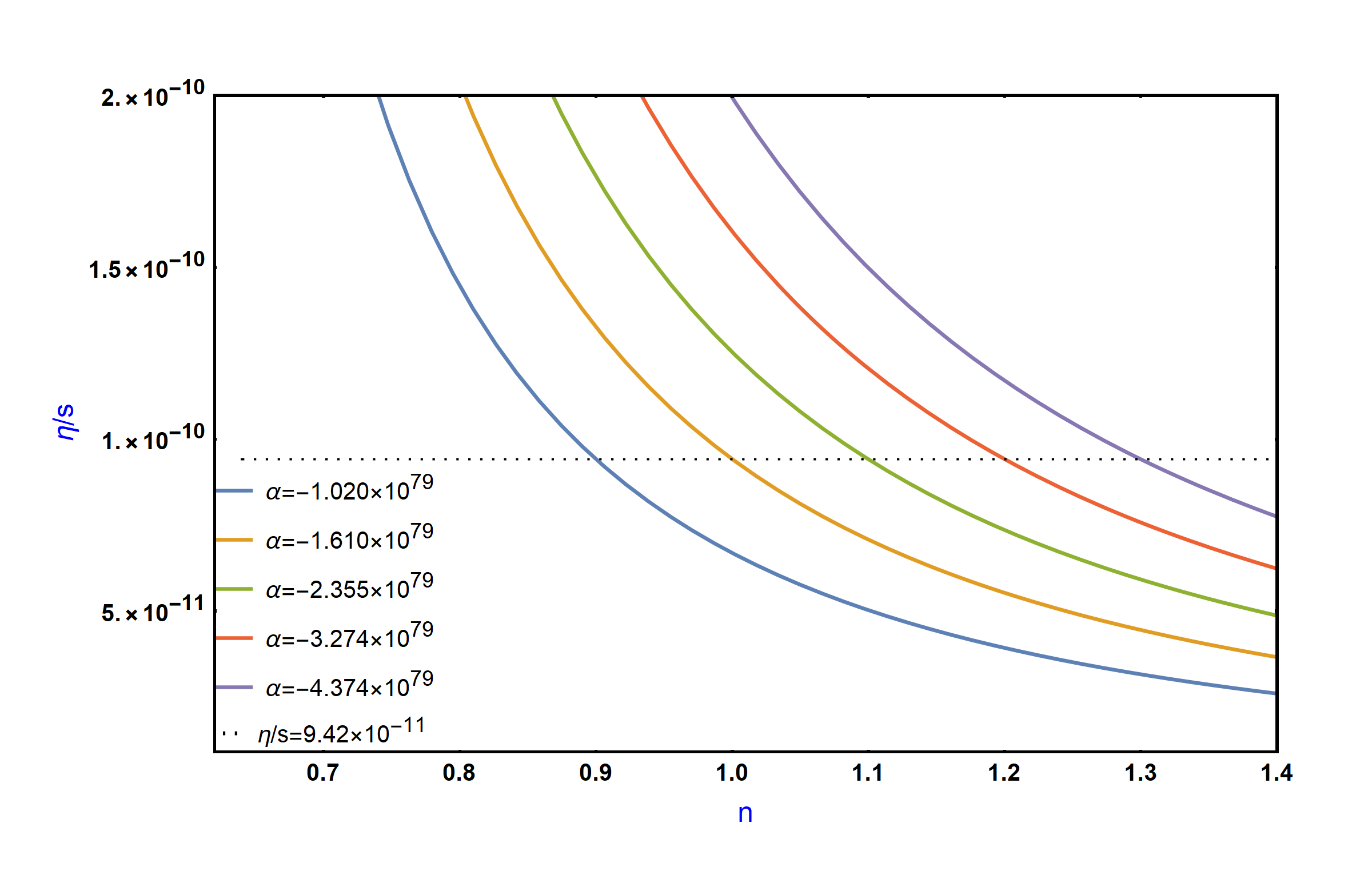}
  \caption{\justifying Plot of \(\frac{\eta}{s}\) versus \(n\) for different values of \(\alpha\) with \(g_b \simeq 1\), \(M_*=1\times10^{12} GeV\), \(T_D=2\times10^{16} GeV\), \(g_*=106\) and \(\beta=1\).}
  \label{fig:Figure8}
\end{subfigure}
\hfill
\begin{subfigure}[b]{0.45\textwidth}
  \centering
   \hspace*{-0.5cm} 
  \includegraphics[width=1.10\linewidth, trim=0 0 1cm 1cm, clip]{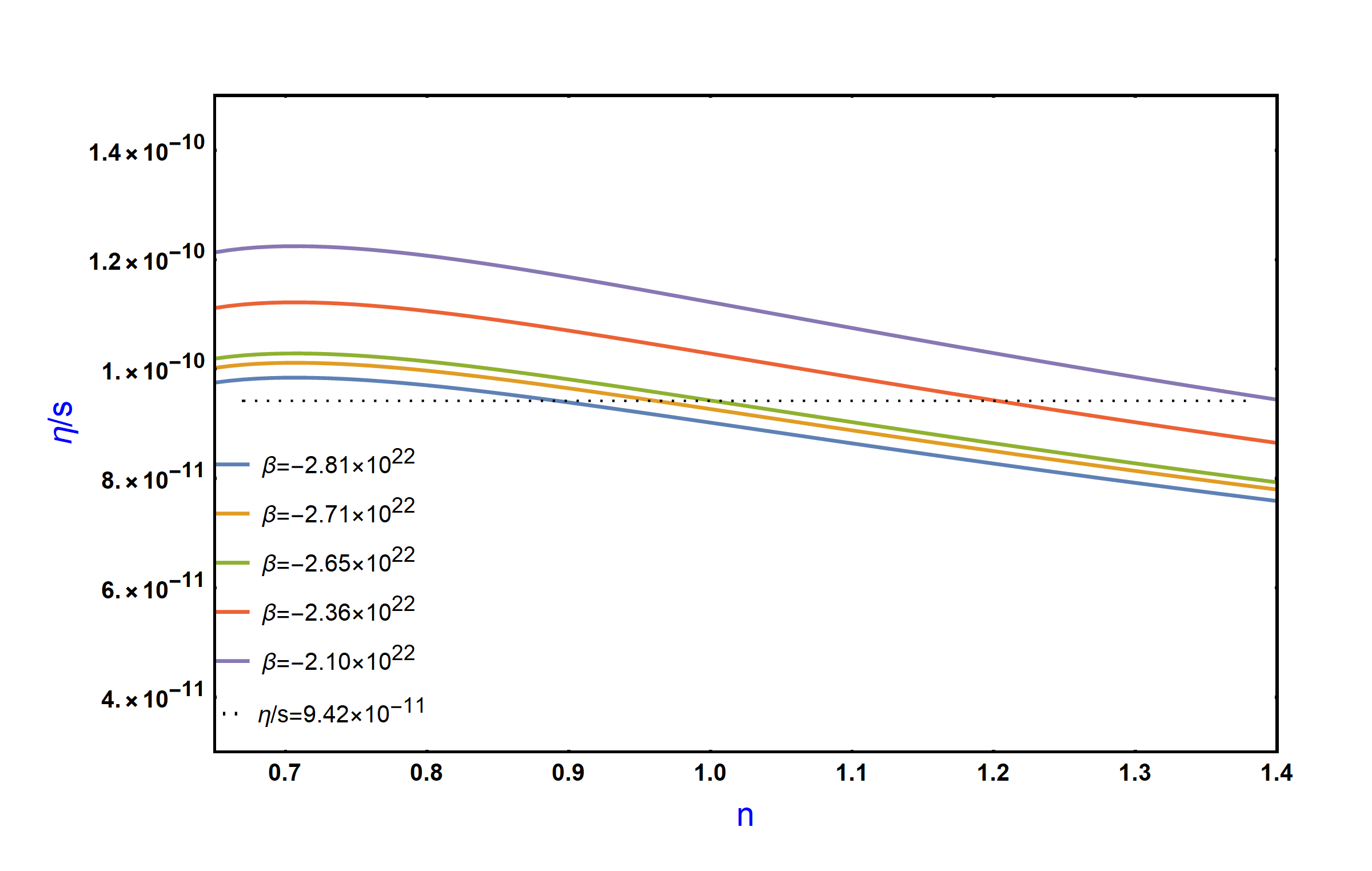}
  \caption{\justifying Plot of \(\frac{\eta}{s}\) versus \(n\) for different values of \(\beta\) with \(g_b \simeq 1\), \(M_*=1\times10^{12} GeV\), \(T_D=2\times10^{16} GeV\), \(g_*=106\) and \(\alpha = -5 \times 10^{96}\).}
  \label{fig:Figure9}
\end{subfigure}

\caption{Plot of \(\frac{\eta}{s}\) as a function of \(\alpha\), \(\beta\) and \(n\) for Case II (\(m=2\)) of Model I.}
\label{fig:fig_case2_model1}
\end{figure*}

Inserting \(f^{\prime}\) and \(f^{\prime\prime}\) in the first motion eqn. \eqref{eq:friedmann_equation1}, the energy density cab be derived analytically as: 
\begin{equation}
 -\frac{2 \beta  k^2 n^4 (14 n+11) (1-2 n)^4+2 \alpha  k^2 (10 n-83) t^8+n^3 (2 n-1)^3 (8 n-7) t^2}{4 k^2 n^2 (2 n-1)^3 t^4}
 \label{eq:energy_density_model1_case2}
\end{equation}
The expression of the energy density in eqn. \eqref{eq:energy_density_model1_case2} upon equating with the energy density in eqn. \eqref{eq:energy_density}, the decoupling time \((t_D)\) can be expressed as a function of the  decoupling temperature \((T_D)\):

\begin{equation}
t_D = \sqrt{\frac{1}{2} \sqrt{2 M+N+P-Q}+\frac{1}{2} \sqrt{M-N-P}}
\label{eq:tD_for_model1_case2}
\end{equation}
where
\begin{align*}
Q &= \left(\frac{(n^3 (-1 + 2 n)^3 (-7 + 8 n)}{(-83 + 10 n) P \alpha}\right), 
\hspace{0.5cm} P = \left(\frac{\sqrt[3]{\sqrt{G^2-4 F^3}+G}}{6 \sqrt[3]{2} \alpha  (10 n-83)}\right), \hspace{0.5cm}
M = \left(\frac{A}{3 \alpha  (10 n-83)}\right), \\
N &= \left(\frac{F}{(3\times2^{2/3} \left(G + \sqrt{-4 F^3 + G^2}\right)^{1/3} (-83 + 10 n) \alpha)}\right), 
\hspace{0.3cm} F=\left(A^2+48 B\right),
\hspace{0.3cm} G=\left(2 A^3-288 A B-54 C\right),\\
A &=\left(2/15 g_* n^2 \pi^2 T_D^4 - 4/5 g_* n^3 \pi^2 T_D^4 + 
 8/5 g_* n^4 \pi^2 T_D^4 - 16/15 g_* n^5 \pi^2 T_D^4\right),\\
B &=\left( (1-2 n)^4 n^4 (10 n-83) (14 n+11)\alpha  \beta \right),
\hspace{2.5cm} C = \left(\alpha  n^6 (2 n-1)^6 (8 n-7)^2 (10 n-83)\right).
\end{align*}
Using eqns. \eqref{eq:ricci_scalar},\eqref{eq:general_baryontoentropyratio} and \eqref{eq:tD_for_model1_case2}, we can derive the expression for the \(\frac{\eta}{s}\) ratio as follows: 
\begin{eqnarray}
\frac{\eta}{s}=  \frac{15 g_b }{4 \pi ^2 g_* M_*^2 T_D}\left(\frac{2 (n-1) n}{\mathcal{S}_1^{3/2}}+\frac{2 n^2}{\mathcal{S}_2^{3/2}}\right)
\label{eq:ratio_model1_case2}
\end{eqnarray}
where
\begin{align*}
\mathcal{S}_1 &= \left( \frac{1}{2} \sqrt{-\mathcal{J} - \mathcal{K} + \mathcal{H}} + \frac{\sqrt{\mathcal{L}}}{2} \right),
\hspace{4cm} \mathcal{S}_2 = \left( \frac{1}{2} \sqrt{-\mathcal{J} + \mathcal{K} + \mathcal{H}} + \frac{\sqrt{\mathcal{L}}}{2} \right), \\
\mathcal{L}  &= \left( \mathcal{J} + \mathcal{K} - \frac{\mathcal{D}}{\alpha (10n - 83) \sqrt{-\mathcal{J} - \mathcal{K} + \mathcal{H}}} + 2 \mathcal{H} \right), \hspace{0.9cm} \mathcal{H} = \left( \frac{\mathcal{A}}{3 \alpha (10n - 83)} \right), \\
\mathcal{J}  &= \left( \frac{\mathcal{F}}{3 \cdot 2^{2/3} \alpha (10n - 83) \sqrt[3]{\mathcal{G} + \sqrt{\mathcal{G}^2 - 4 \mathcal{F}^3}}} \right),
\hspace{1.8cm} \mathcal{K} = \left(\frac{\sqrt[3]{\mathcal{G}+\sqrt{\mathcal{G}^2-4 \mathcal{F}^3}}}{6 \sqrt[3]{2} \alpha  (10 n-83)}\right),\\
\mathcal{F} &= \left(\mathcal{A}^2+48 \mathcal{B}\right),
\hspace{1cm} \mathcal{G} = \left(2 \mathcal{A}^3-288 \mathcal{A} \mathcal{B}-54 \mathcal{C}\right),
\hspace{1.2cm} \mathcal{D} = \left(n^3 (2 n-1)^3 (8 n-7)\right),\\
\mathcal{A} &=\left(-\frac{1}{15} 16 \pi ^2 g_* n^5 T_D^4+\frac{8}{5} \pi ^2 g_* n^4 T_D^4-\frac{4}{5} \pi ^2 g_* n^3 T_D^4+\frac{2}{15} \pi ^2 g_* n^2 T_D^4\right),\\
\mathcal{B} &=\left(\alpha  \beta  (1-2 n)^4 n^4 (10 n-83) (14 n+11)\right),
\hspace{2.5cm} \mathcal{C} = \left(\alpha  n^6 (2 n-1)^6 (8 n-7)^2 (10 n-83)\right).
\end{align*}
The eqn. \eqref{eq:ratio_model1_case2} predicts a non-vanishing baryon-to-entropy ratio for \(n\neq\frac{1}{2}\), contrasting with GR. The eqn. \eqref{eq:ratio_model1_case2} diverge at \(n=\frac{1}{2}\). This divergence makes \(n=\frac{1}{2}\) an excluded case in our analysis. The values of the constant parameters of eqn. \eqref{eq:ratio_model1_case2} are \(g_b \simeq 1\), \(g_*=106\), \(M_*=1\times10^{12} GeV\) and \(T_D=2\times10^{16} GeV\). In fig. \ref{fig:Figure6}, \(\frac{\eta}{s}\) is plotted against \(\alpha\) for five distinct values of \(n\), maintaining \(\beta\) at a constant value: \(\beta=1\) and in fig. \ref{fig:Figure7}, \(\frac{\eta}{s}\) is plotted as a function of \(\beta\) for various values of \(n\) taking \(\alpha\) fixed at: \(\alpha=-5\times10^{96}\). \autoref{tab:tableB} presents a list of values of  \(n\), \(\beta\) and \(\frac{\eta}{s}\), for which the theoretically derived values of the asymmetry ratio closely match the observed asymmetry ratio obtained from cosmological data. Finally, \(\frac{\eta}{s}\) is plotted against n for various \(\alpha\) values in fig. \ref{fig:Figure8}and \(\beta\) values in fig. \ref{fig:Figure9}.

\begin{table}[htb]
  \caption{For Case II (\(m = 2\)) of Model I, by fixing \(\alpha = -5 \times 10^{96}\), the table below presents a list of \(n\) and \(\beta\) values that yield a matter imbalance ratio closely aligned with the observed asymmetry ratio.}
    \begin{tabular}{c@{\hspace{1cm}}c@{\hspace{1cm}}c}
    \hline \hline
    $n$ & $\beta$ & $\frac{\eta}{s}$ \\
    \hline
    0.8 & $-2.71 \times 10^{22}$ & $9.971 \times 10^{-11}$ \\
    1.0 & $-2.64 \times 10^{22}$ & $9.453 \times 10^{-11}$ \\
    1.2 & $-2.36 \times 10^{22}$ & $9.429 \times 10^{-11}$ \\
    \hline \hline
    \end{tabular}
    \label{tab:tableB}
\end{table}

\begin{figure*}[htbp]
\centering
\begin{subfigure}[b]{0.48\columnwidth}
  \centering
  \includegraphics[width=\linewidth]{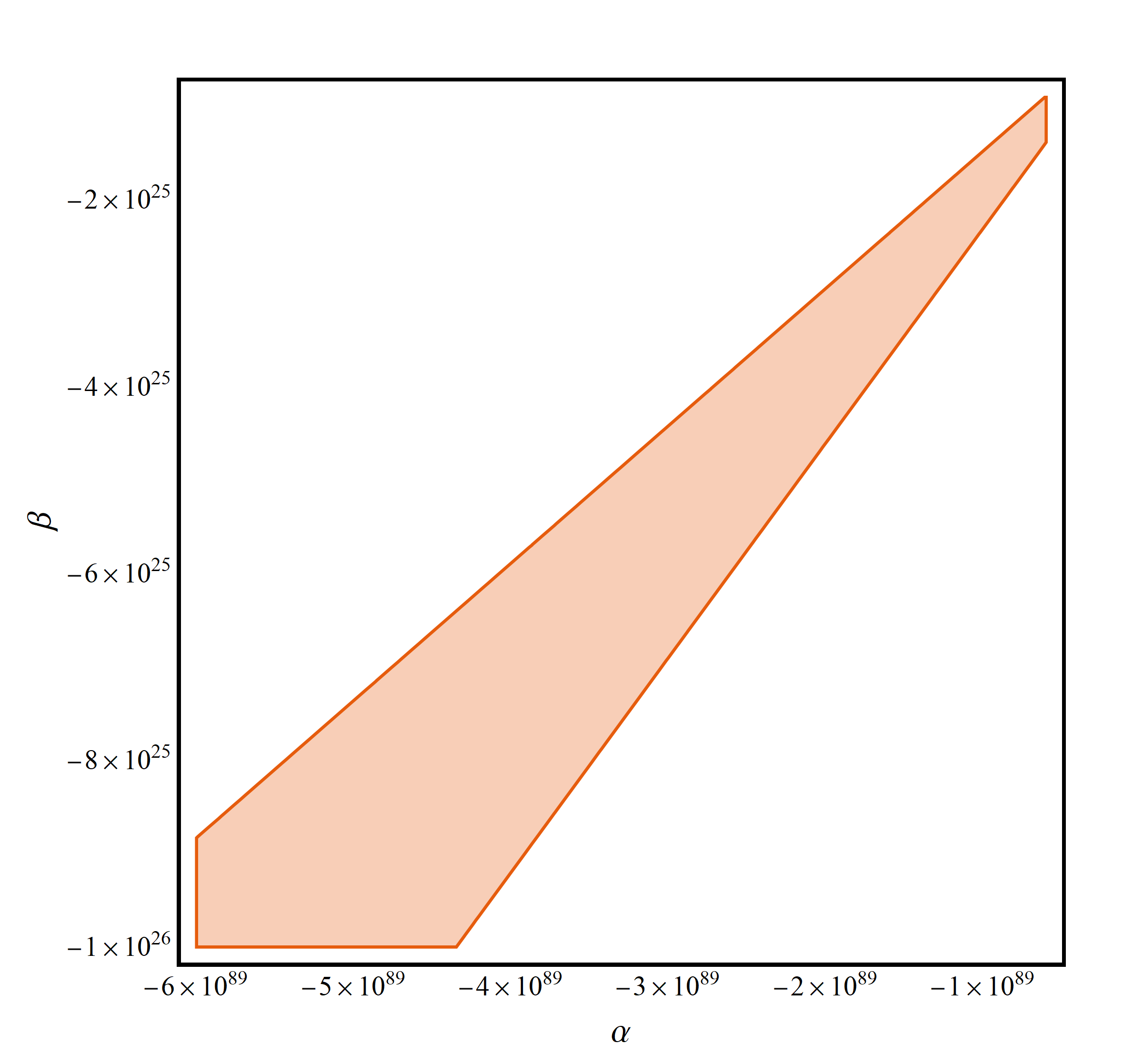}
  \caption{\justifying Plot of the parameter space of \(\alpha\) and \(\beta\) for the Case I (\(m=1\)) of Model I, for which \(8\times10^{-11}\leq \frac{\eta}{s} \leq 2\times10^{-10}\). Here, we have, \(g_b \simeq 1\), \(M_*=1\times10^{12} GeV\), \(T_D=2\times10^{16} GeV\), \(g_*=106\) and \(n=0.6\).}
  \label{fig:Figure5}
\end{subfigure}%
\hfill
\begin{subfigure}[b]{0.48\columnwidth}
  \centering
  \includegraphics[width=\linewidth]{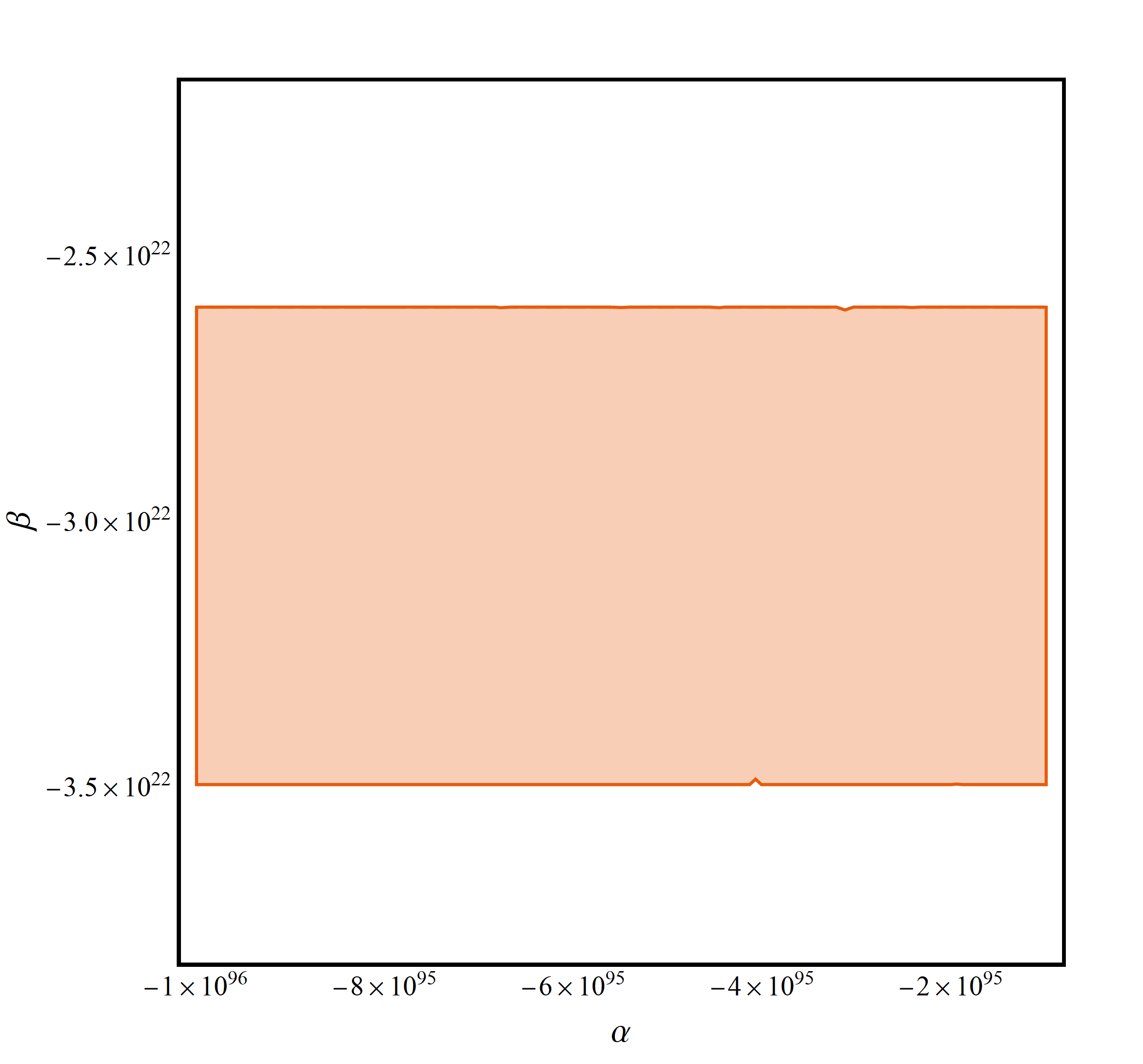}
  \caption{\justifying Plot in parameter space of \(\alpha\) and \(\beta\) for the Case II (\(m=2\)) of Model I, for which \(8\times10^{-11}\leq \frac{\eta}{s} \leq 2\times10^{-10}\). Here, we have, \(g_b \simeq 1\), \(M_*=1\times10^{12} GeV\), \(T_D=2\times10^{16} GeV\), \(g_*=106\) and \(n=0.6\).}
  \label{fig:Figure10}
\end{subfigure}
\caption{Plot of the parameter space of \(\alpha\) and \(\beta\) for the two cases of Model I}
\label{fig:fig_model1_parameter_space}
\end{figure*}

\subsection{Model II}

In Model II, we utilize the extended gravity functional form, which is of the following form \cite{nojiri2008modified}:
\begin{eqnarray}
f(R)=\alpha R^{\gamma} +\beta R^{\delta}
\end{eqnarray}
\hspace{1cm} here, \(\alpha\) and \(\beta\) are positive constants, and \(\gamma\) and \(\delta\) are positive integers satisfying the condition \(\gamma > \delta\).

In the higher curvature epoch of the early universe, the term: \(\alpha R^{\gamma}\) dominates the rapid expansion \cite{nojiri2003modified,odintsov2016singular} and as cosmic expansion reduces the curvature scale, the term: \(-\beta R^{\delta}\) becomes dynamically significant, effectively performing the function of dark energy in driving the observed accelerated expansion at late times. This model combines an early inflationary epoch with a subsequent DE phase in a single \(f(R)\) framework that naturally transits between high and low-curvature eras \cite{nojiri2006modified}, all within a minimal modified-gravity setting.

In order to solve the eqns. \eqref{eq:friedmann_equation1} and \eqref{eq:friedmann_equation2} analytically for model II $f(R)$, we have to specify the numerical values of \(\gamma\) and \(\delta\). Here, we consider two pairs of values for (\(\gamma\),\(\delta\)), they are: (2,1) and (4,1).

\subsubsection{Case I}
For Case I, we consider the integer value for the pair (\(\gamma\), \(\delta\)) as $(2,1)$. The $f(R)$ takes the form: 

\begin{eqnarray}
f(R)=\alpha R^2+\beta R
\end{eqnarray}

We evaluate the energy density expression by substituting \(f^{\prime}\) and \(f^{\prime\prime}\) in eqn. \eqref{eq:friedmann_equation1}, which yields the following:

\begin{eqnarray}
\frac{n \left(\alpha  n (11-4 n (7 n+2))+\beta  (7-8 n) t^2\right)}{2 t^4}
\label{eq:energy_density_Model2_case1}
\end{eqnarray}

Equating the eqns. \eqref{eq:friedmann_equation1} and \eqref{eq:energy_density_Model2_case1}, the decoupling time \((t_D)\) can be calculated as a function of the decoupling temperature \((T_D)\) as follows: 

{\small
\begin{equation}
t_D=\sqrt{-\frac{60 \beta  n^2}{\pi ^2 g_* T^4}+\frac{\sqrt{\left(120 \beta  n^2-105 \beta  n\right)^2-4 \pi ^2 g_* T^4 \left(420 \alpha  n^4+120 \alpha  n^3-165 \alpha  n^2\right)}}{2 \pi ^2 g_* T^4}+\frac{105 \beta  n}{2 \pi ^2 g_* T^4}}
\label{eq:tD_for_model2_case1}
\end{equation}
}

\(\frac{\eta}{s}\) can be evaluated using eqns. \eqref{eq:ricci_scalar},\eqref{eq:general_baryontoentropyratio} and \eqref{eq:tD_for_model2_case1} and the expression is given as follows: 

\begin{equation}
\frac{\eta}{s} = \frac{15 g_b}{2 g_* M_*^2 \pi^2 T_D} \left(\frac{n (2 n-1)}{S^{3/2}} \right)
\label{eq:ratio_model2_case1}
\end{equation}
where
\begin{align*}
S &=\left(\frac{\sqrt{B^2-4 \pi ^2 A g_* T_D^4}}{2 \pi ^2 g_* T_D^4}+C\right),
\hspace{2cm} C= \left(\frac{105 n \beta}{(2 g_* \pi^2 T_D^4} - \frac{60 n^2 \beta}{
 g_* \pi^2 T_D^4}\right),\\
A &=\left(420 \alpha  n^4+120 \alpha  n^3-165 \alpha  n^2\right), 
\hspace{1.6cm} B= \left(120 \beta  n^2-105 \beta  n\right)
\end{align*}

\begin{figure*}[htbp]
\centering
\begin{subfigure}[b]{0.48\columnwidth}
  \centering
   \hspace*{-1cm} 
  \includegraphics[width=1.10\linewidth, trim=0 0 1cm 1cm, clip]{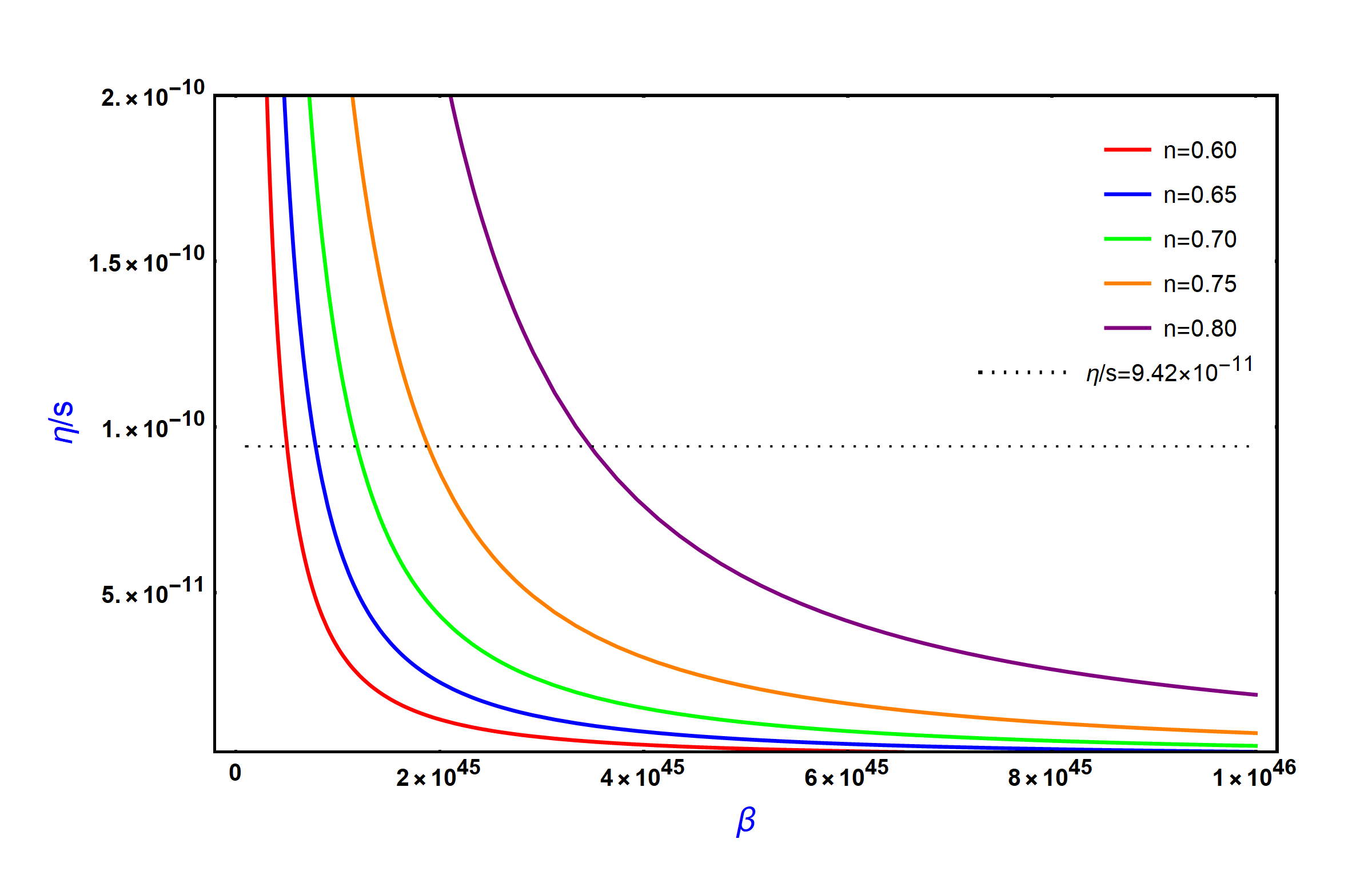}
  \caption{\justifying Plot of \(\frac{\eta}{s}\) versus \(\beta\) for different values of \(n\) with \(g_b \simeq 1\), \(M_*=1\times10^{12} GeV\), \(T_D=2\times10^{16} GeV\), \(g_*=106\) and \(\alpha=1\).}
  \label{fig:Figure11}
\end{subfigure}%
\hfill
\begin{subfigure}[b]{0.48\columnwidth}
  \centering
   \hspace*{-0.5cm} 
  \includegraphics[width=1.10\linewidth, trim=0 0 1cm 1cm, clip]{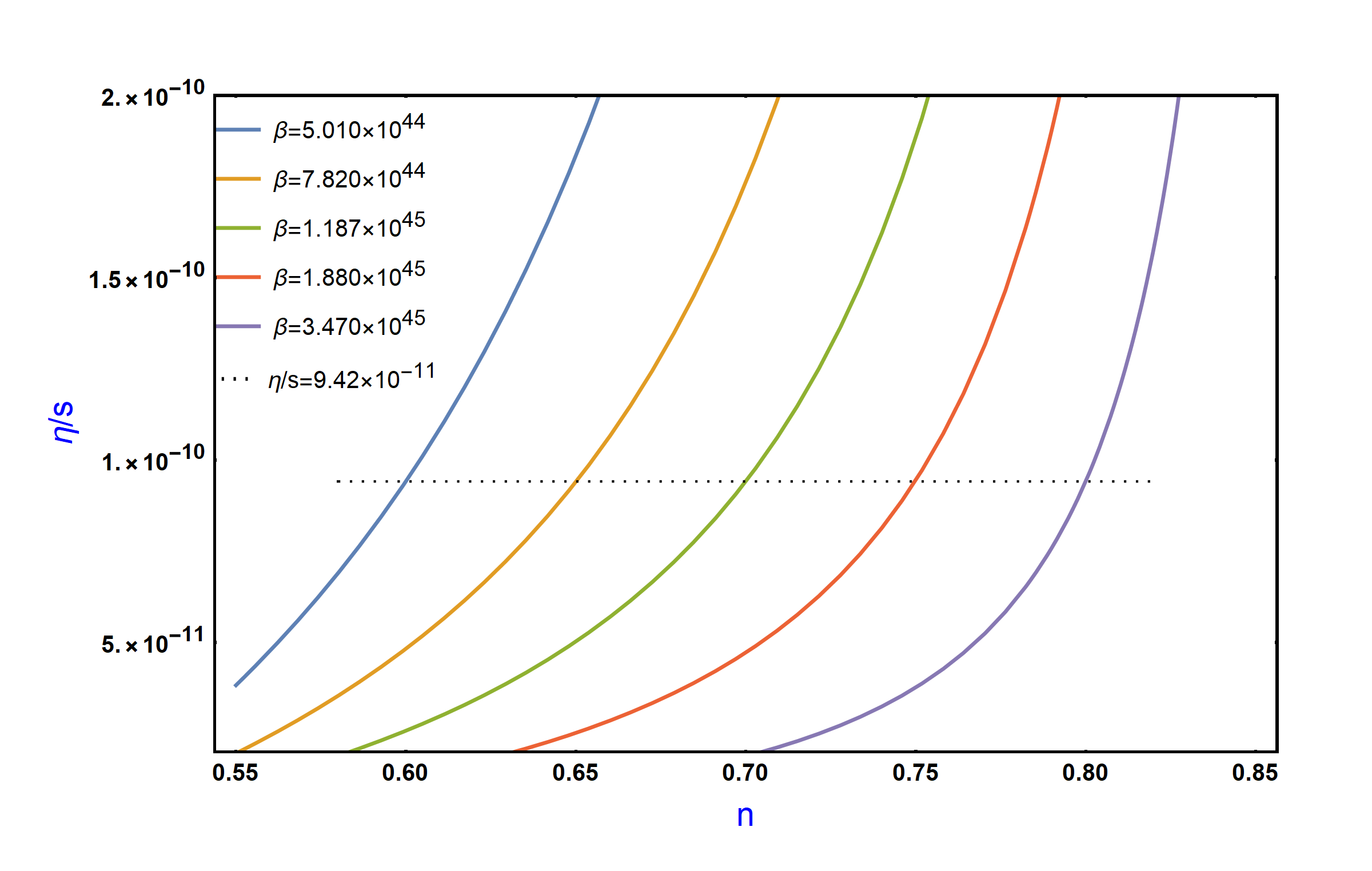}
  \caption{\justifying Plot of \(\frac{\eta}{s}\) versus \(n\) for different values of \(\beta\) with \(g_b \simeq 1\), \(M_*=1\times10^{12} GeV\), \(T_D=2\times10^{16} GeV\), \(g_*=106\) and \(\alpha=1\).}
  \label{fig:Figure12}
\end{subfigure}
\caption{Plot of \(\frac{\eta}{s}\) as a function of \(\beta\) and \(n\) for Case I (\(\ \gamma=2 , \delta=1 \)) of Model II.}
\label{fig:fig_case1_model2}
\end{figure*}

Unlike in GR, \(\frac{\eta}{s}\) derived from eqn. \eqref{eq:ratio_model2_case1} yields a non-zero value when \(n\neq\frac{1}{2}\). The parameter choice, \(n=\frac{1}{2}\), produces a singularity in eqn. \eqref{eq:ratio_model2_case1}. Therefore, the value \(n=\frac{1}{2}\) is trivial in our model. Substituting \(g_b \simeq 1\), \(g_*=106\), \(M_*=1\times10^{12} GeV\), \(T_D=2\times10^{16} GeV\) and \(n=0.7\), with model parameters \(\alpha=1\) and \(\beta=1.19\times10^{45}\), into eqn. \eqref{eq:ratio_model2_case1}, the resulting \(\frac{\eta}{s}\) becomes \(9.42\times10^{-11}\), which is in high agreement with the observational constraint. \autoref{tab:tableC} outlines a set of selected values of \(n\), \(\beta\) and \(\frac{\eta}{s}\) that result in theoretical estimates of the asymmetry ratio that are in strong agreement with those derived from cosmological observations. Fig. \ref{fig:Figure11} and fig. \ref{fig:Figure12} present \(\frac{\eta}{s}\) as a function of \(\beta\) and \(n\), respectively.

\begin{table}[htb]
  \caption{For Case I (\(\ \gamma=2 , \delta=1 \)) of Model II, by fixing \(\alpha = 1\), we have a table of values of \(n\) and \(\beta\) for which the BnER, \(\frac{\eta}{s}\), agrees well with the observed value of the asymmetry ratio.}
  \begin{tabular}{c@{\hspace{1cm}}c@{\hspace{1cm}}c}
    \hline \hline
    $n$ & $\beta$ & $\frac{\eta}{s}$ \\
    \hline
    0.6 & $5.01 \times 10^{44}$ & $9.423 \times 10^{-11}$ \\
    0.7 & $1.19 \times 10^{45}$ & $9.425 \times 10^{-11}$ \\
    0.8 & $3.45 \times 10^{45}$ & $9.430 \times 10^{-11}$ \\
    \hline \hline
  \end{tabular}
  \label{tab:tableC}
\end{table}

\subsubsection{Case II}

For Case II, we use \(f(R)\) model of the form: 

\begin{eqnarray}
f(R)=\alpha  R^4+\beta  R
\label{eq:fR_form_model2_case2}
\end{eqnarray}

Using eqn. \eqref{eq:scalefactor} in eqn. \eqref{eq:energy_density} along with the eqn. \eqref{eq:fR_form_model2_case2} and its derivatives, the mathematical expression of the energy density can be obtained as: 

\begin{eqnarray}
\frac{(1-2 n)^2 n^3 \left(\beta  (-(20 n+53)) t^2-\alpha  n (2 n-1) (26 n+119)\right)}{2 t^8}
\label{eq:energy_density_model2_case2}
\end{eqnarray}

Equating eqns. \eqref{eq:friedmann_equation1} and \eqref{eq:energy_density_model2_case2}, we arrive at the decoupling time \((t_D)\) as a function of the decoupling temperature \((T_D)\), as follows: 

\begin{eqnarray}
t_D &= \left(\sqrt{\frac{1}{2} \sqrt{\frac{3375 C^2}{\sqrt{G}}+\frac{225 C^2}{2}-\frac{4 \sqrt[3]{\frac{2}{3}} D}{\sqrt[3]{F}}-\frac{5 \sqrt[3]{\frac{3}{2}} \sqrt[3]{F}}{\pi ^2 g_* T_D^4}}+\frac{15 C}{4}+\frac{\sqrt{G}}{8}}\right)
\label{eq:tD_model2-case2}
\end{eqnarray}

\hspace{1cm} where

\begin{align*}
G &=\left(900 C^2+\frac{64 \sqrt[3]{\frac{2}{3}} D}{\sqrt[3]{F}}+\frac{40\ 2^{2/3} \sqrt[3]{3} \sqrt[3]{F}}{\pi ^2 g_* T_D^4}\right), \hspace{2.1cm} F=\left(\sqrt{3} \sqrt{27 A^2+B}+9 A\right),\\
A &=\left(\alpha  \beta ^2 (7-8 n)^2 n^6 (2 n-1)^3 (26 n+119)\right),
\hspace{2.2cm} C=\left(\frac{\beta  (7-8 n) n}{\pi ^2 g_* T_D^4}\right),\\
B &=\left(-\frac{256 \pi ^6 \alpha ^3 g_*^3 n^{12} (2 n-1)^9 (26 n+119)^3 T^{12}}{3375}\right), 
\hspace{1cm} D = \left(\alpha  n^4 (2 n-1)^3 (26 n+119)\right)
\end{align*}

Upon using the expression of decoupling time, the BnER defined in eqn. \eqref{eq:general_baryontoentropyratio} takes the following form: 
\begin{equation}
\frac{\eta}{s}=\frac{15 g_b }{2 \pi ^2 g_* M_*^2  T_D} \frac{n (2 n-1)}{\mathcal{S}^{3/2}}
\label{eq:ratio_model2_case2}
\end{equation}
where
\begin{align*}
\mathcal{S} &=\left(\frac{15 \mathcal{C}}{4}+\frac{\sqrt{\mathcal{G}}}{8}+\frac{1}{2} \sqrt{\frac{3375 \mathcal{C}^3}{\sqrt{\mathcal{G}}}+\frac{225 \mathcal{C}^2}{2}-\frac{4 \sqrt[3]{\frac{2}{3}} \mathcal{D}}{\sqrt[3]{\mathcal{F}}}-\frac{5 \sqrt[3]{\frac{3}{2}} \sqrt[3]{\mathcal{F}}}{\pi ^2 \text{gstr} T^4}}\right),\\
\mathcal{G} &= \left(900 \mathcal{C}^2+\frac{64 \sqrt[3]{\frac{2}{3}} \mathcal{D}}{\sqrt[3]{\mathcal{F}}}+\frac{40\ 2^{2/3} \sqrt[3]{3} \sqrt[3]{\mathcal{F}}}{\pi ^2 g_* T_D^4}\right),
\hspace{2.4cm} \mathcal{F}=\left(\sqrt{3} \sqrt{27 \mathcal{A}^2+\mathcal{B}}+9 \mathcal{A}\right),\\
\mathcal{A} &=\left(\alpha  \beta ^2 (7-8 n)^2 n^6 (2 n-1)^3 (26 n+119)\right),
\hspace{2.5cm} \mathcal{C}=\left(\frac{\beta  (7-8 n) n}{\pi ^2 g_* T_D^4}\right),\\
\mathcal{B} &=\left(\frac{-256 g_*^3 n^{12} (-1 + 2 n)^9 (119 + 
    26 n)^3 \pi^6 T_D^{12} \alpha^3}{3375}\right),
\hspace{1cm} \mathcal{D}=\left(n^4 (-1 + 2 n)^3 (119 + 26 n) \alpha\right)
\end{align*}

\begin{figure*}[htbp]
\centering
\begin{subfigure}[b]{0.48\columnwidth}
  \centering
   \hspace*{-1cm} 
  \includegraphics[width=1.10\linewidth, trim=0 0 1cm 1cm, clip]{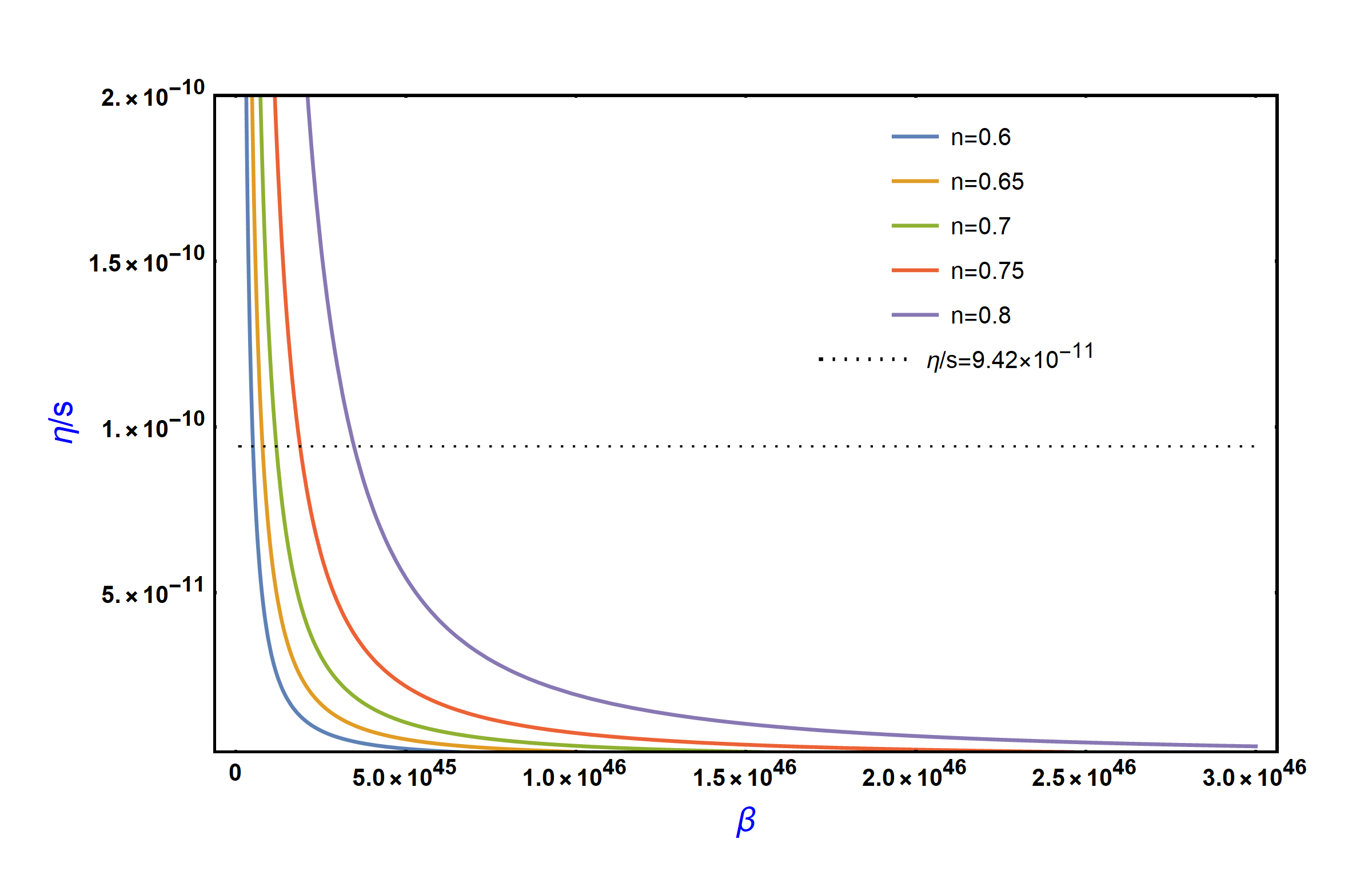}
  \caption{\justifying Plot of \(\frac{\eta}{s}\) versus \(\beta\) for different values of \(n\) with \(g_b \simeq 1\), \(M_*=1\times10^{12} GeV\), \(T_D=2\times10^{16} GeV\), \(g_*=106\) and \(\alpha=1\times10^{-25}\).}
  \label{fig:Figure14}
\end{subfigure}%
\hfill
\begin{subfigure}[b]{0.48\columnwidth}
  \centering
   \hspace*{-0.5cm} 
  \includegraphics[width=1.10\linewidth, trim=0 0 1cm 1cm, clip]{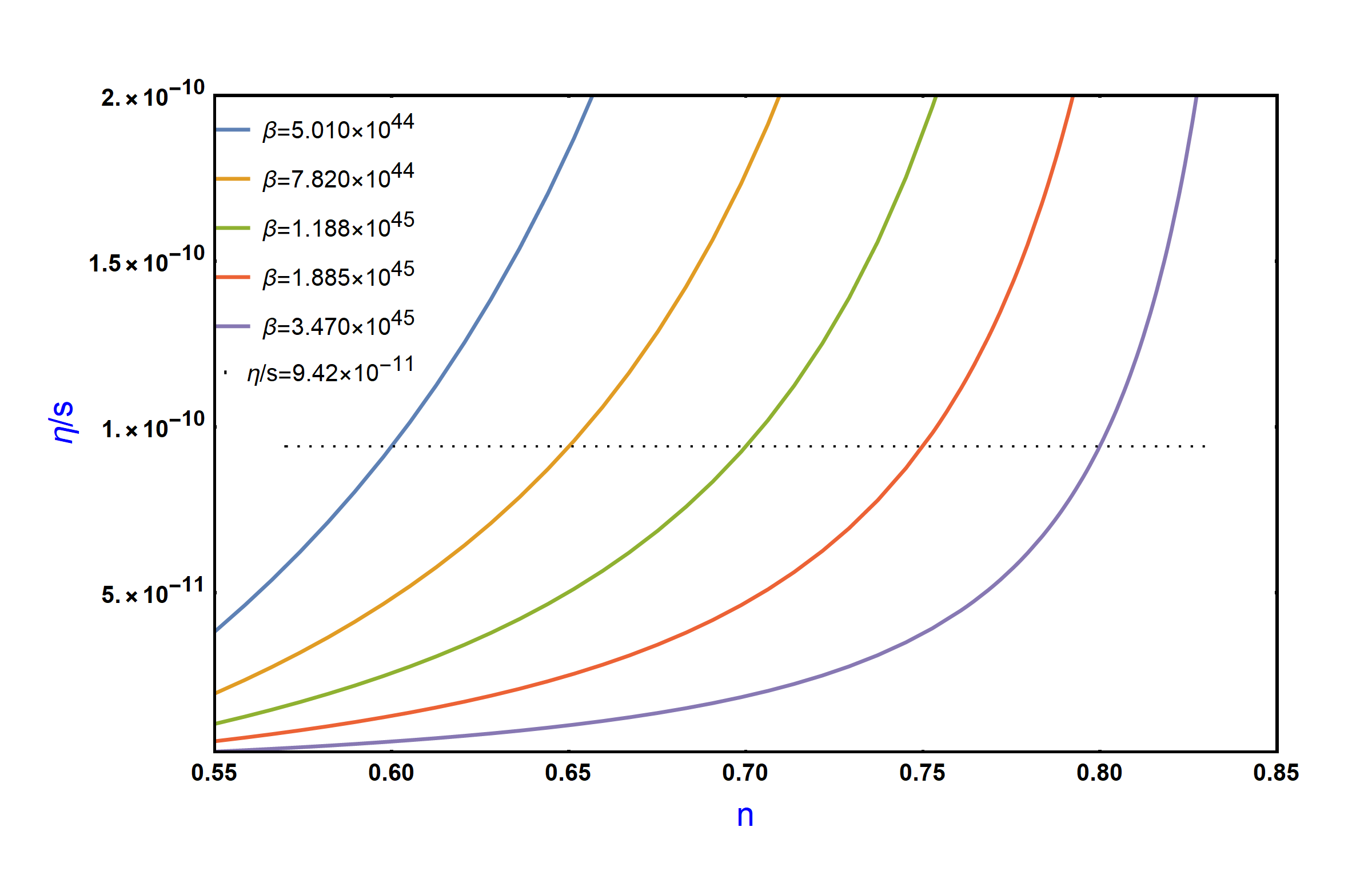}
  \caption{\justifying Plot of \(\frac{\eta}{s}\) versus \(n\) for different values of \(\beta\) with \(g_b \simeq 1\), \(M_*=1\times10^{12} GeV\), \(T_D=2\times10^{16} GeV\), \(g_*=106\) and \(\alpha=1\times10^{-25}\).}
  \label{fig:Figure15}
\end{subfigure}
\caption{Plot of \(\frac{\eta}{s}\) as a function of \(\beta\) and \(n\) for Case II (\(\ \gamma=4 , \delta=1 \)) of Model II.}
\label{fig_case2_model2}
\end{figure*}

While GR gives a null result, our model in eqn. \eqref{eq:ratio_model2_case2} generates a finite baryon asymmetry for \(n\neq\frac{1}{2}\). The special case \(n=\frac{1}{2}\) is pathological, as it induces complex infinity. Due to the anomaly, \(n=\frac{1}{2}\) does not work in our model. By choosing \(g_b\), \(g_*\), \(M_*\) and \(T_D\) as in the earlier case, with the model parameters: \(\alpha=10^{-25}\), \(\beta=1.19\times10^{45}\) and \(n=0.7\), the final BnER becomes equal to \(9.415\times10^{-11}\), which is in excellent consistency with observational data. \autoref{tab:tableD} presents specific values of \(n\), \(\beta\), and \(\frac{\eta}{s}\) in tabulated format that yield theoretical predictions for the asymmetry ratio closely aligned with observational data from cosmological studies. In fig. \ref{fig:Figure14} and fig. \ref{fig:Figure15}, \(\frac{\eta}{s}\) is plotted as a function of \(\beta\) and \(n\) respectively.

\begin{table}[htb]
  \caption{For Case II (\(\ \gamma=4 , \delta=1 \)) of Model II, by fixing \(\alpha = 10^{-25}\), the table below provides a set of values of \(n\) and \(\beta\) for which the BnER \((\frac{\eta}{s})\) agrees well with the observed value of the asymmetry ratio.}
  \begin{tabular}{c@{\hspace{1cm}}c@{\hspace{1cm}}c}
    \hline \hline
    $n$ & $\beta$ & $\frac{\eta}{s}$ \\
    \hline
    0.6 & $2.75 \times 10^{44}$ & $9.425 \times 10^{-11}$ \\
    0.7 & $1.19 \times 10^{45}$ & $9.415 \times 10^{-11}$ \\
    0.8 & $3.47 \times 10^{45}$ & $9.433 \times 10^{-11}$ \\
    \hline \hline
  \end{tabular}
  \label{tab:tableD}
\end{table}

\begin{figure*}[htbp]
\centering
\begin{subfigure}[b]{0.48\columnwidth}
  \centering
  \includegraphics[width=\linewidth]{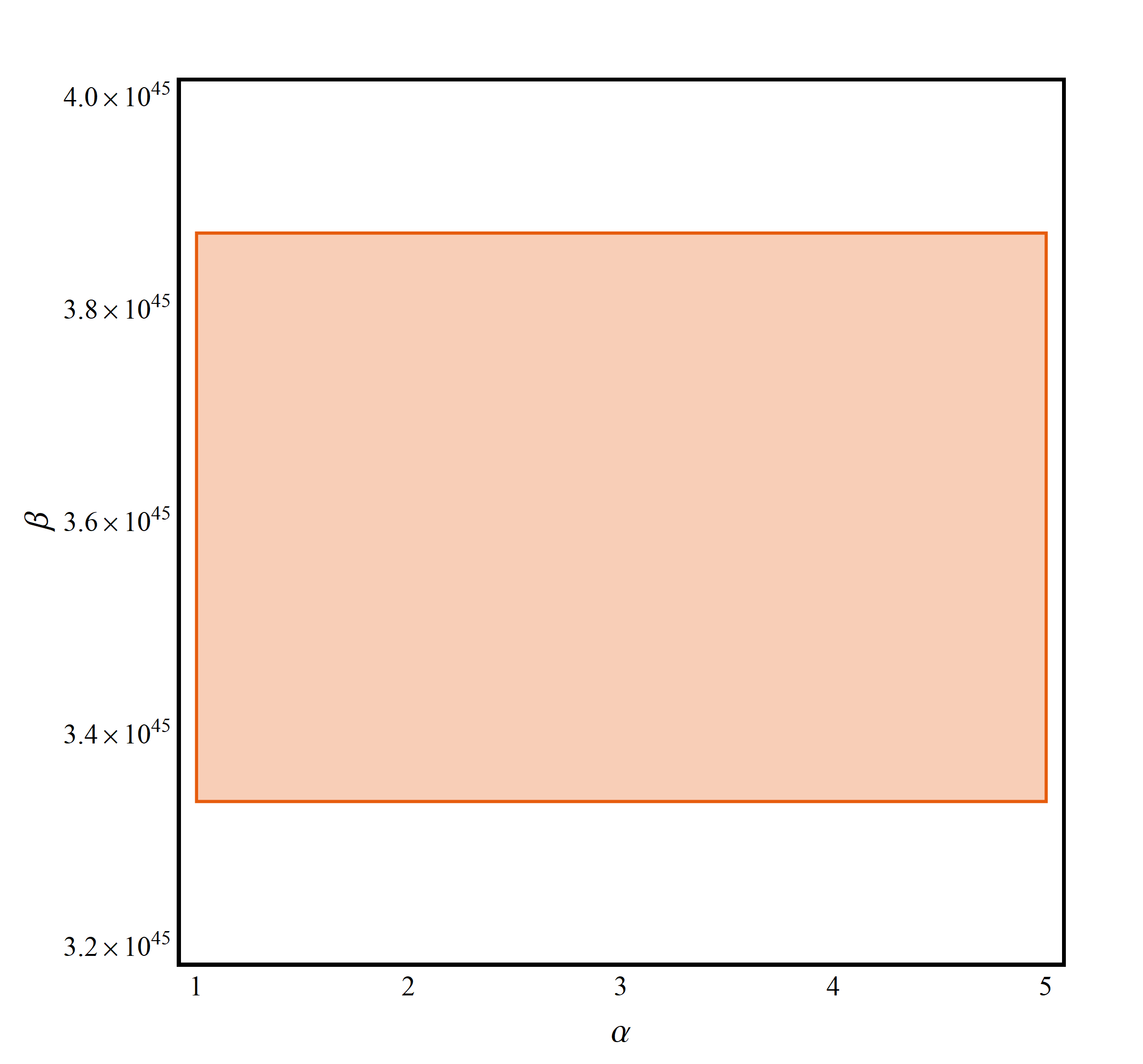}
  \caption{\justifying Plot of the parameter space of \(\alpha\) and \(\beta\) for the Case I (\(\ \gamma=2 , \delta=1 \)) of Model II, for which \(8\times10^{-11}\leq \frac{\eta}{s} \leq 2\times10^{-10}\). Here, we have, \(g_b \simeq 1\), \(M_*=1\times10^{12} GeV\), \(T_D=2\times10^{16} GeV\), \(g_*=106\) and \(n=0.6\).}
  \label{fig:Figure13}
\end{subfigure}%
\hfill
\begin{subfigure}[b]{0.48\columnwidth}
  \centering
  \includegraphics[width=\linewidth]{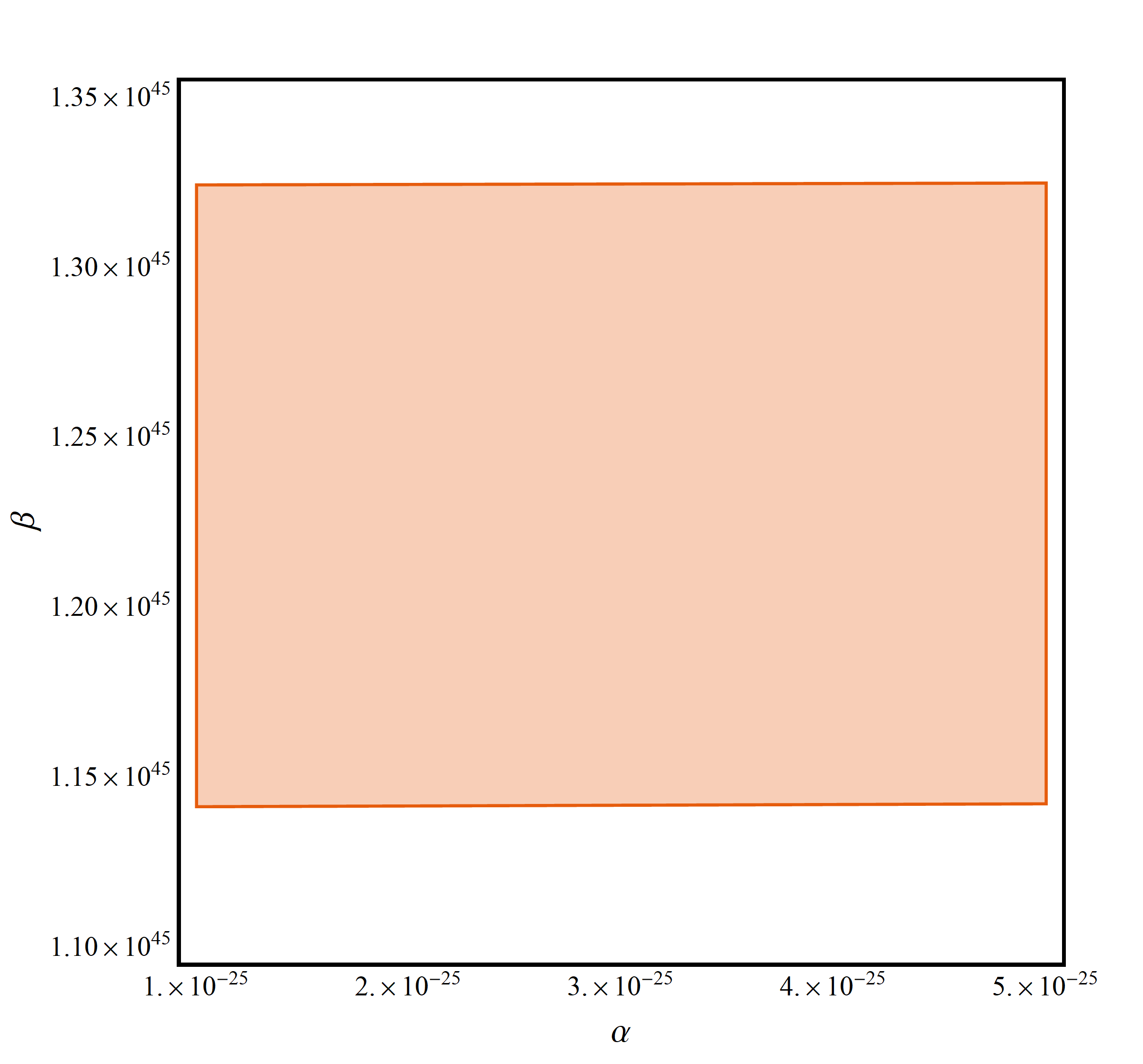}
  \caption{\justifying Plot of the parameter space of \(\alpha\) and \(\beta\) for the Case II (\(\ \gamma=4 , \delta=1 \)) of Model II, for which \(8\times10^{-11}\leq \frac{\eta}{s} \leq 2\times10^{-10}\). Here, we have, \(g_b \simeq 1\), \(M_*=1\times10^{12} GeV\), \(T_D=2\times10^{16} GeV\), \(g_*=106\) and \(n=0.6\).}
  \label{fig:Figure16}
\end{subfigure}
\caption{Plot of the parameter space of \(\alpha\) and \(\beta\) for the two cases of Model II}
\label{fig_model2_parameter_space}
\end{figure*}

\subsection{Model III}
Model III is a rational form $f(R)$ gravity model proposed in \cite{nojiri2008modified}, having the functional form:
\begin{eqnarray}
f(R)=\frac{\alpha R^{2m}-\beta R^m}{1+\gamma R^m}
\end{eqnarray}

\hspace{1cm} here, \(m\) is a positive integer and \(\alpha\), \(\beta\), and \(\gamma\) are constant model parameters having positive values. \\

The functional form of Model III \(f(R)\) is designed to ensure a smooth transition from an early inflationary period (where the effects of higher-order curvature terms like \(R^2\) prevail) to a late-time acceleration epoch that aligns with \(\Lambda\)CDM cosmological observations \cite{nojiri2006modified}, naturally bridging the early inflationary epoch and the later dark energy driven era \cite{nojiri2008modified}.

Instead of working in a general functional form of $f(R)$ for Model III, we will specify the $f(R)$ form by giving a particular positive integer value to m that allows us to effectively solve eqns. \eqref{eq:friedmann_equation1} and \eqref{eq:friedmann_equation2}  using analytical methods. Here, we consider \(m=1\).

For $m = 1$, the $f(R)$ model takes the form as follows: 

\begin{eqnarray}
f(R)=\frac{\left(\alpha R^{2}-\beta R\right)}{\left(1+\gamma R\right)}
\label{eq:fR_model3}
\end{eqnarray}

The analytical expression for the energy density can be obtained from the temporal component of the field equation \eqref{eq:friedmann_equation1} by inserting eqns. \eqref{eq:scalefactor} \& \eqref{eq:fR_model3} as follows:

\begin{eqnarray}
&& [n (-\gamma ^2 n^2 (2 n-1)^3 \left(\alpha  n (8 n-7)-\beta  t^2\right)+\beta  (8 n-7) t^6+\alpha  n (11-4 n (7 n+2)) t^4-2 \gamma  n (2 n-1) t^2 (\alpha  n (2 n-1) (11 n-10)
 \nonumber\\&& +\beta  (16-5 n) t^2)]/2 t^2 \left(\gamma  n (2 n-1)+t^2\right)^3
 \label{eq:energy_density_model3}
\end{eqnarray}
Setting the expression of the energy density in eqn. \eqref{eq:energy_density_model3} equal to that in eqn. \eqref{eq:energy_density},the decoupling time \((t_D)\) can be derived as a function of the decoupling temperature \((T_D)\):
\begin{eqnarray}
t_D &= \left(\sqrt{\frac{1}{2} \sqrt{-\mathcal{J}+\frac{\mathcal{K}}{4 \sqrt{\mathcal{J}+\mathcal{H}}}+\mathcal{M}}+\frac{\sqrt{\mathcal{J}+\mathcal{H}}}{2}+\mathcal{L}}\right)
\label{eq:tD_model3}
\end{eqnarray}
where
\begin{align*}
\mathcal{H} &=\left(\frac{225 \mathcal{A}^2}{4 \pi ^4 g_*^2 T_D^8}+\frac{10 \mathcal{B}}{\pi ^2 g_* T_D^4}\right), 
\hspace{3.2cm} \mathcal{J} = \left(\frac{5 \sqrt[3]{\mathcal{G}+\sqrt{\mathcal{G}^2-4 \mathcal{F}^3}}}{\sqrt[3]{2} \pi ^2 g_* T_D^4}+\frac{5 \sqrt[3]{2} \mathcal{F}}{\pi ^2 g_* T_D^4 \sqrt[3]{\mathcal{G}+\sqrt{\mathcal{G}^2-4 \mathcal{F}^3}}}\right),\\
\mathcal{K} &=\left(\frac{3375 \mathcal{A}^3}{\pi ^6 g_*^3 T_D^{12}}+\frac{900 \mathcal{A} \mathcal{B}}{\pi ^4 g_*^2 T_D^8}+\frac{120 \mathcal{D}}{\pi ^2 g_* T_D^4}\right),
\hspace{1.3cm} \mathcal{L}=\left(\frac{15 \mathcal{A}}{4 \pi ^2 g_* T_D^4}\right),
\hspace{1cm} \mathcal{M}=\left(\frac{225 \mathcal{A}^2}{2 \pi ^4 g_*^2 T_D^8}+\frac{20 \mathcal{B}}{\pi ^2 g_* T_D^4}\right),\\
\mathcal{F} &=\left(-3 \mathcal{A} \mathcal{D}-\frac{1}{5} 4 \pi ^2 \mathcal{C} g_* T_D^4+\mathcal{B}^2\right),
\hspace{2.2cm} \mathcal{G} =\left(-27 \mathcal{A}^2 \mathcal{C}+9 \mathcal{A} \mathcal{D} \mathcal{B}+\frac{9}{5} \pi ^2 \mathcal{D}^2 g_* T_D^4-\frac{24}{5} \pi ^2 \mathcal{C} g_* \mathcal{B} T_D^4-2 \mathcal{B}^3\right),\\
\mathcal{A} &=\left(-\frac{1}{5} 2 \pi ^2 \gamma  g_* n^2 T_D^4+\frac{1}{5} \pi ^2 \gamma  g_* n T_D^4+8 \beta  n^2-7 \beta  n\right),\\
\mathcal{B} &=\left(-\frac{4}{5} \pi ^2 \gamma ^2 g_* n^4 T_D^4+\frac{4}{5} \pi ^2 \gamma ^2 g_* n^3 T_D^4-\frac{1}{5} \pi ^2 \gamma ^2 g_* n^2 T_D^4-28 \alpha  n^4+20 \beta  \gamma  n^4-8 \alpha  n^3-74 \beta  \gamma  n^3+11 \alpha  n^2+32 \beta  \gamma  n^2\right),\\
\mathcal{C} &=\left(-64 \alpha  \gamma ^2 n^8+152 \alpha  \gamma ^2 n^7-132 \alpha  \gamma ^2 n^6+50 \alpha  \gamma ^2 n^5-7 \alpha  \gamma ^2 n^4\right),\\
\mathcal{D} &=(-\frac{8}{15} \pi ^2 \gamma ^3 g_* n^6 T_D^4+\frac{4}{5} \pi ^2 \gamma ^3 g_* n^5 T_D^4-\frac{2}{5} \pi ^2 \gamma ^3 g_* n^4 T_D^4+\frac{1}{15} \pi ^2 \gamma ^3 g_* n^3 T_D^4-88 \alpha  \gamma  n^6+8 \beta  \gamma ^2 n^6+168 \alpha  \gamma  n^5 \\& -12 \beta  \gamma ^2 n^5-102 \alpha  \gamma  n^4+6 \beta  \gamma ^2 n^4+20 \alpha  \gamma  n^3-\beta  \gamma ^2 n^3).
\end{align*}

Using eqn. \eqref{eq:baryon-number-to-entropy}, the expression for the BnER can be obtained by inserting eqns. \eqref{eq:ricci_scalar} \& \eqref{eq:tD_model3}:

{\large
\begin{eqnarray}
\frac{\eta}{s} &=\frac{15 g_b }{2 \pi ^2 g_* M_*^2 T_D} \left(\frac{n (2 n-1)}{S^{3/2}}\right)
\label{eq:ratio_model3}
\end{eqnarray}
}

where

\begin{align*}
S &=\left(\frac{15 A}{4 \pi ^2 g_* T_D^4}+\frac{1}{2} \sqrt{\frac{K}{4 \sqrt{L+M}}+N-P}+\frac{\sqrt{L+M}}{2}\right),\\
P &=\left(\frac{5 \sqrt[3]{\sqrt{J^2-4 Q^3}+J}}{\sqrt[3]{2} \pi ^2 g_* T_D^4}+\frac{5 \sqrt[3]{2} Q}{\pi ^2 g_* T_D^4 \sqrt[3]{\sqrt{J^2-4 Q^3}+J}}\right),
\hspace{1cm} K=\left(\frac{3375 A^3}{\pi ^6 g_*^3 T_D^{12}}+\frac{900 A B}{\pi ^4 g_*^2 T_D^8}+\frac{120 F}{\pi ^2 g_* T_D^4}\right),\\
L &=\left(\frac{225 A^2}{4 \pi ^4 g_*^2 T_D^8}+\frac{10 B}{\pi ^2 g_* T_D^4}\right),
\hspace{2.8cm} M=\left(\frac{5 \sqrt[3]{\sqrt{J^2-4 Q^3}+J}}{\sqrt[3]{2} \pi ^2 g_* T_D^4}+\frac{5 \sqrt[3]{2} Q}{\pi ^2 g_* T_D^4 \sqrt[3]{\sqrt{J^2-4 Q^3}+J}}\right),\\
N &=\left(\frac{225 A^2}{2 \pi ^4 g_*^2 T_D^8}+\frac{20 B}{\pi ^2 g_* T_D^4}\right),
\hspace{6cm} Q=\left(-3 A F+B^2-\frac{1}{5} 4 \pi ^2 g_* G T_D^4\right),\\
J &=\left(-27 A^2 G+9 A B F-2 B^3-\frac{24}{5} \pi ^2 B g_* G T_D^4+\frac{9}{5} \pi ^2 F^2 g_* T_D^4\right),\\
A &=\left(-\frac{1}{5} 2 \pi ^2 \gamma  g_* n^2 T_D^4+\frac{1}{5} \pi ^2 \gamma  g_* n T_D^4+8 \beta  n^2-7 \beta  n\right),\\
B &=\left(-\frac{4}{5} \pi ^2 \gamma ^2 g_* n^4 T_D^4+\frac{4}{5} \pi ^2 \gamma ^2 g_* n^3 T_D^4-\frac{1}{5} \pi ^2 \gamma ^2 g_* n^2 T_D^4-28 \alpha  n^4+20 \beta  \gamma  n^4-8 \alpha  n^3-74 \beta  \gamma  n^3+11 \alpha  n^2+32 \beta  \gamma  n^2\right),\\
C &=\left(-128 \alpha  \gamma ^2 n^8+304 \alpha  \gamma ^2 n^7-264 \alpha  \gamma ^2 n^6+100 \alpha  \gamma ^2 n^5-14 \alpha  \gamma ^2 n^4\right),\\
F &=(-\frac{8}{15} \pi ^2 \gamma ^3 g_* n^6 T_D^4+\frac{4}{5} \pi ^2 \gamma ^3 g_* n^5 T_D^4-\frac{2}{5} \pi ^2 \gamma ^3 g_* n^4 T_D^4+\frac{1}{15} \pi ^2 \gamma ^3 g_* n^3 T_D^4-88 \alpha  \gamma  n^6+8 \beta  \gamma ^2 n^6+168 \alpha  \gamma  n^5- \\& 12 \beta  \gamma ^2 n^5-102 \alpha  \gamma  n^4+6 \beta  \gamma ^2 n^4+20 \alpha  \gamma  n^3-\beta  \gamma ^2 n^3),\\
G &=\left(-64 \alpha  \gamma ^2 n^8+152 \alpha  \gamma ^2 n^7-132 \alpha  \gamma ^2 n^6+50 \alpha  \gamma ^2 n^5-7 \alpha  \gamma ^2 n^4\right).
\end{align*}

\begin{figure}[htbp]
  \centering
  \begin{subfigure}[b]{0.48\linewidth} 
    \centering
    \hspace*{-1cm} 
    \includegraphics[width=1.10\linewidth, trim=0 0 1cm 1cm, clip]{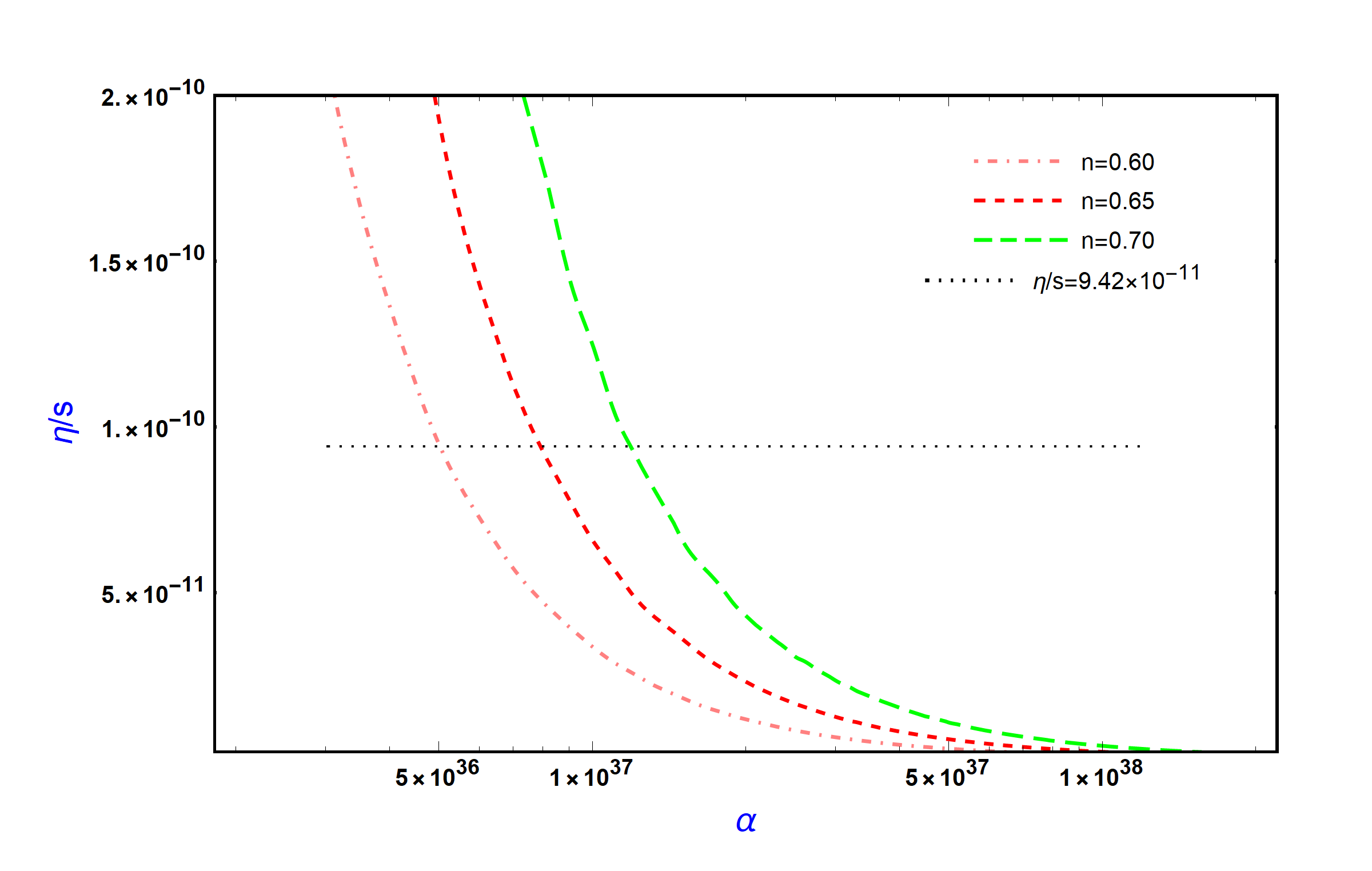} 
    \caption{\justifying Plot of \(\frac{\eta}{s}\) versus \(\alpha\) for different values of \(n\) with \(g_b \simeq 1\), \(M_*=1\times10^{12} GeV\), \(T_D=2\times10^{16} GeV\), \(g_*=106\), \(\beta=1\) and \(\gamma=1\times10^{-8}\).}
    \label{fig:Figure17}
  \end{subfigure}
  \hfill
  \begin{subfigure}[b]{0.48\linewidth}
    \centering
    \hspace*{-0.5cm} 
    \includegraphics[width=1.10\linewidth, trim=0 0 1cm 1cm, clip]{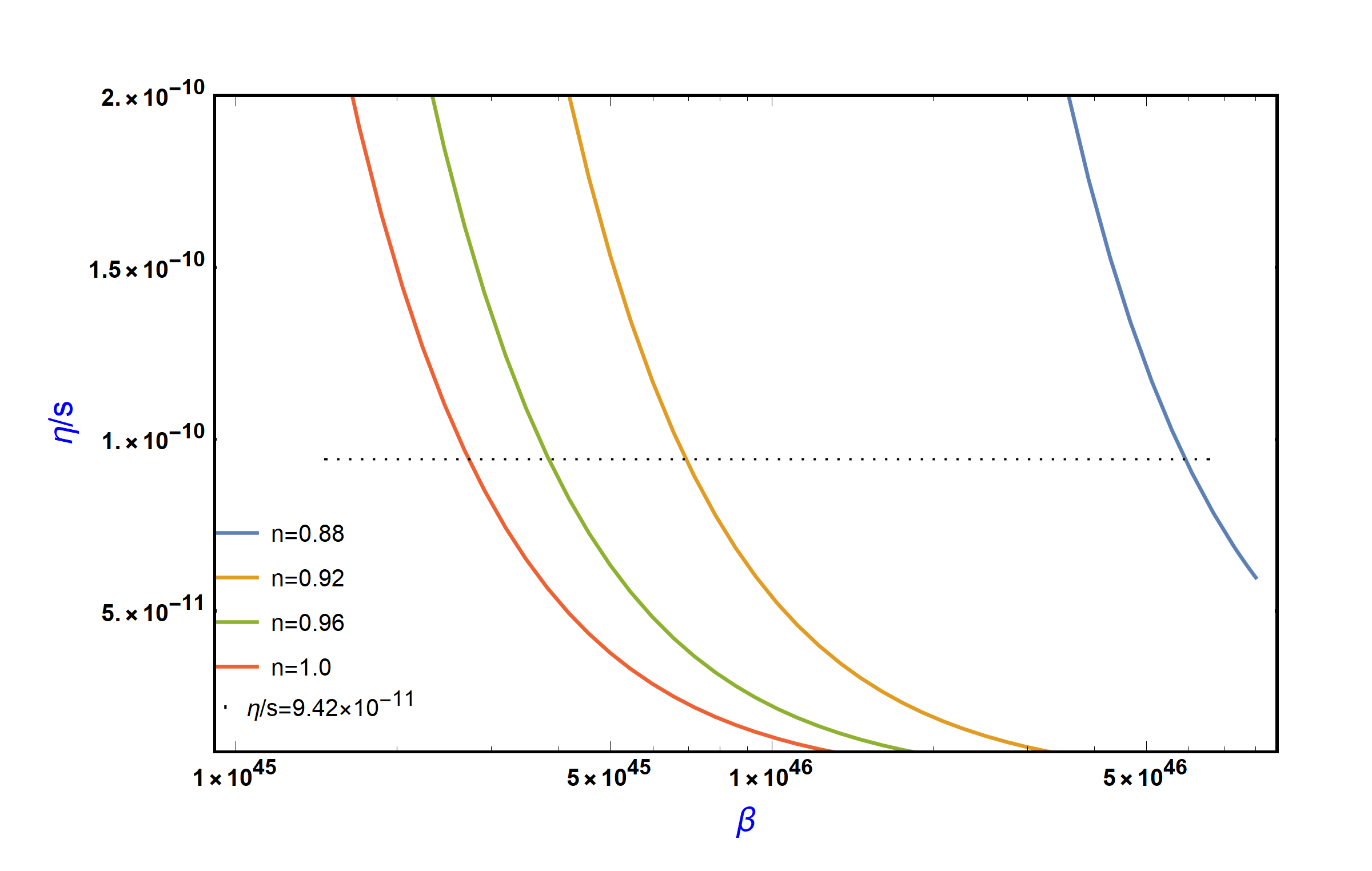}
    \caption{\justifying Plot of \(\frac{\eta}{s}\) versus \(\beta\) for different values of \(n\) with \(g_b \simeq 1\), \(M_*=1\times10^{12} GeV\), \(T_D=2\times10^{16} GeV\), \(g_*=106\), \(\alpha=1\) and \(\gamma=1\times10^{-48}\).}
    \label{fig:Figure18}
  \end{subfigure}

\vspace{0.5cm}

\begin{subfigure}[b]{0.6\linewidth}
  \centering
  \includegraphics[width=\linewidth, trim=0 0 1cm 1cm, clip]{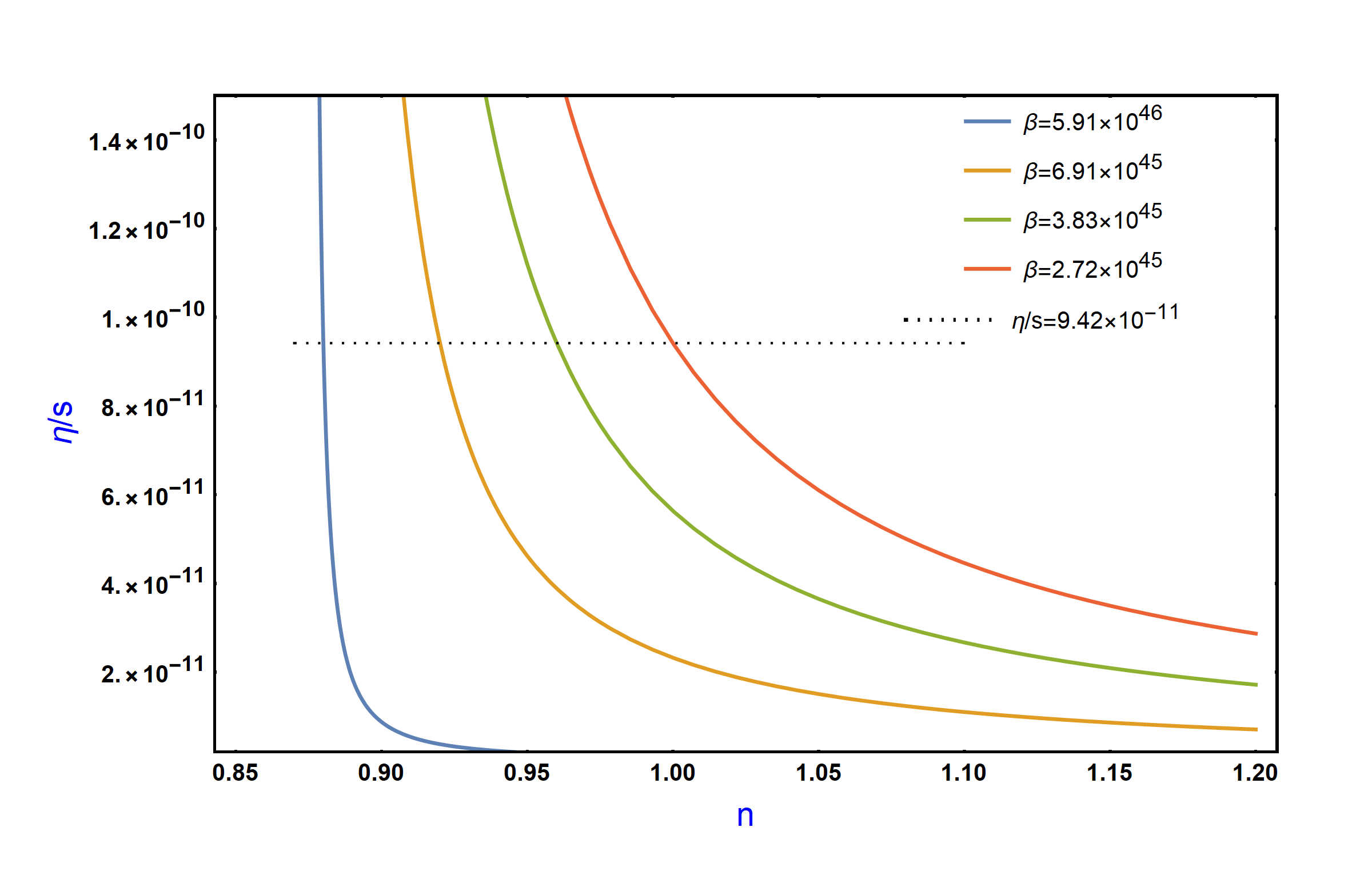}
  \caption{\justifying Plot of \(\frac{\eta}{s}\) versus \(n\) for different values of \(\beta\) with \(g_b \simeq 1\), \(M_*=1\times10^{12} GeV\), \(T_D=2\times10^{16} GeV\), \(g_*=106\), \(\alpha=1\) and \(\gamma=1\times10^{-48}\).}
  \label{fig:Figure19}
\end{subfigure}

\caption{Plot of \(\frac{\eta}{s}\) as a function of \(\alpha\), \(\beta\) and \(n\) for the Case, \(m=1 \), of Model III.}
\label{fig:threeimagegroup}
\end{figure}

The baryon number density normalized by entropy, obtained from eqn. \eqref{eq:ratio_model3}, vanishes in GR but remains non-zero in our framework provided \(n\neq\frac{1}{2}\). The value \(n=\frac{1}{2}\) is unphysical for our model as it leads to an undefined ratio. Substituting the constant parameter \(g_b\), \(g_*\), \(M_*\) and \(T_D\) as before with model parameters \(\beta=1\) and \(\gamma=1\times10^{-8}\), we can plot the baryon imbalance ratio \(\frac{\eta}{s}\) as a function of \(\alpha\) as shown in fig. \ref{fig:Figure17}. The values of \(n\), \(\alpha\), and \(\frac{\eta}{s}\)  listed in \autoref{tab:tableE} predict asymmetry ratios that align well with the observational range. Similarly, by fixing the model parameters at \(\alpha = 1\) and \(\gamma=1\times 10^{-48}\), \(\frac{\eta}{s}\) is plotted as a function of \(\beta\) depicted in fig. \ref{fig:Figure18}. \autoref{tab:tableF} displays selected values of \(n\) , \(\beta\) and \(\frac{\eta}{s}\) that produce theoretical results of the asymmetry ratio in close agreement with the observed cosmological data. In fig. \ref{fig:Figure19}, \(\frac{\eta}{s}\) is plotted against \(n\) for four distinct values of \(\beta\). The dotted horizontal line represents the astronomically measured value of BnER.

\begin{table}[htb]
  \caption{For the Case, m=1, of the Model III, where the parameters are set to \(\beta=1\) and \(\gamma=1\times10^{-8}\), the table provides various combinations of \(n\) and \(\alpha\) that result in a baryon asymmetry ratio matching observational constraint.}
  \begin{tabular}{c@{\hspace{1cm}}c@{\hspace{1cm}}c}
    \hline \hline
    $n$ & $\alpha$ & $\frac{\eta}{s}$ \\
    \hline
    0.6 & $5.0 \times 10^{36}$ & $9.37 \times 10^{-11}$ \\
    0.7 & $1.2 \times 10^{37}$ & $9.58 \times 10^{-11}$ \\
    0.8 & $3.5 \times 10^{37}$ & $9.64 \times 10^{-11}$ \\
    \hline \hline
  \end{tabular}
  \label{tab:tableE}
\end{table}

\begin{table}[htb]
  \caption{For the Case, m=1, of the Model III, with the parameters fixed at \(\alpha = 1\) and \(\gamma=1\times 10^{-48}\), the table below lists a range of \(n\) and \(\beta\) values that produce baryon imbalance ratio consistent with the observed value.}
  \begin{tabular}{c@{\hspace{1cm}}c@{\hspace{1cm}}c}
    \hline \hline
    $n$ & $\beta$ & $\frac{\eta}{s}$ \\
    \hline
    0.90 & $1.21 \times 10^{46}$ & $9.413 \times 10^{-11}$ \\
    0.94 & $4.90 \times 10^{45}$ & $9.424 \times 10^{-11}$ \\
    0.98 & $3.17 \times 10^{45}$ & $9.425 \times 10^{-11}$ \\
    \hline \hline
  \end{tabular}
  \label{tab:tableF}
\end{table}

\begin{figure}[htbp]
\centering
\includegraphics[width=0.6\columnwidth]{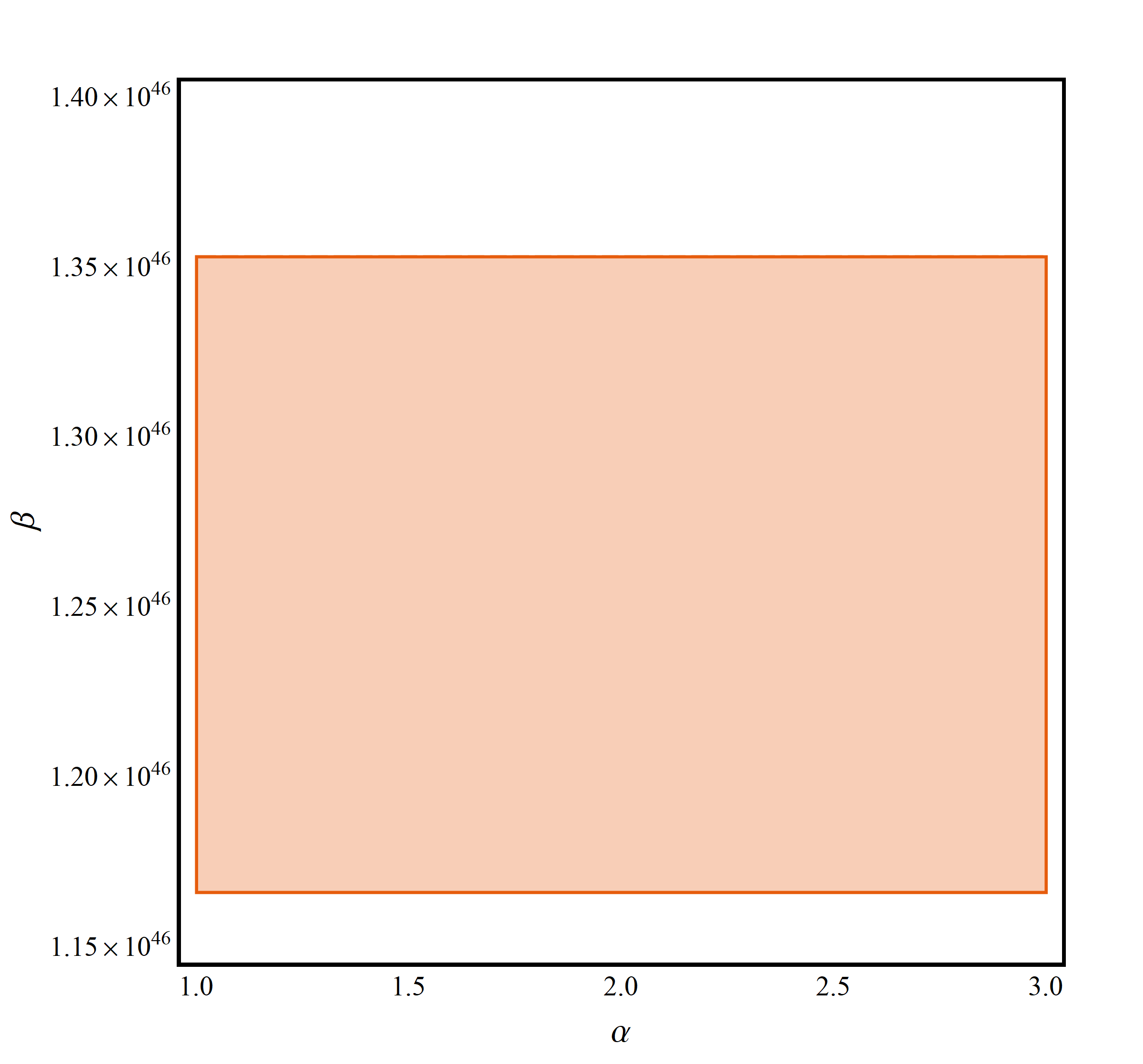}
\caption{Plot in parameter space of \(\alpha\) and \(\beta\) for the Case, \(m=1\), of Model III, for which \(8\times10^{-11}\leq \frac{\eta}{s} \leq 2\times10^{-10}\). Here, we have, \(g_b \simeq 1\), \(M_*=1\times10^{12} GeV\), \(T_D=2\times10^{16} GeV\), \(g_*=106\), \(n=0.9\) and \(\gamma=1\times10^{-48}\).}
\label{fig:Figure20}
\end{figure}

\section{Conclusion}
\label{sec4}

A longstanding puzzle in cosmology is the observed predominance of baryonic matter content over antibaryonic matter content across the cosmos. GB has emerged as one of the potential mechanisms for generating this asymmetry. The idea of GB was originally proposed by Davoudiasl et al. in \cite{davoudiasl2004gravitational}. This work presents a novel investigation of GB within the paradigm of extended $f(R)$ theory, focusing on three distinct and physically motivated models. Although earlier studies such as Lambiase \& Scarpetta \cite{lambiase2006baryogenesis} explored the viability of GB in models like \(f(R)=R^n\), our analysis systematically examines a wider and physically richer set of \(f(R)\) models, incorporating polynomial and rational forms that unify the entire cosmological evolution of the cosmos \cite{nojiri2003modified,nojiri2006modified}. We apply the \(\mathcal{CP}\)-violation coupling \((\partial_\mu R)J^{\mu}\) to the models and derive detailed observational constraints. Through extensive numerical plots and parameter space analysis, we delineate physically acceptable ranges of the model parameters (\(\alpha,\beta,\gamma, n\) etc.) that produce consistent \(\frac{\eta}{s}\) values.  Consequently, our focus lies uniquely on using \(\frac{\eta}{s}\) as a physical probe to obtain an observational constraint for the model parameters, an avenue not explored in those prior works, offering a richer phenomenological landscape advancing the theoretical framework of GB.

This study demonstrates the potential of $f(R)$ gravity frameworks to account for the observed matter-antimatter asymmetry. We examine three distinct $f(R)$ models and analyse various kinds of interactions within each framework, highlighting their viability in explaining the cosmological asymmetry. The model-specific outcomes of this study are demonstrated below:

\begin{enumerate}

    \item \textbf{Model I}: For Model I, we have taken \(f(R)\) of the form: \(f(R)=\frac{-\alpha}{R^m}+\frac{R}{2k^2}+\beta R^2\). We have considered the cases, \(m=1\) and \(m=2\). For both the cases, we have plotted BnER, \(\frac{\eta}{s}\), as a function of \(\alpha\), \(\beta\) and \(n\). 
    
    \begin{itemize}
    
    \item \textbf{Case I}: Here, \(n\) lies within: \(0.55 \leq n \leq 1.0\). The required baryon-to-entropy ratio can be generated only for negative values of the model parameters \(\alpha\) and \(\beta\) as shown in fig. \ref{fig:Figure1} and fig. \ref{fig:Figure2}. The viable upper bound values for the model parameters are: \(\alpha\leq-2\times10^{86}\) and \(\beta\leq-9\times10^{21}\), for the generation of the theoretical results in agreement with that of observed results. There is no such lower bound on values of the model parameters. Fig. \ref{fig:Figure3} and fig. \ref{fig:Figure4} shows the dependence of BnER as a function \(n\) for different values of \(\alpha\) and \(\beta\), respectively.
    
    \item \textbf{Case II}: We have taken the range for \(n\) as: \(0.55 \leq n \leq 1.5\). Like in the earlier case (\(m=1\)), the required BnER can be generated only for non-positive values of the model parameters \(\alpha\) and \(\beta\) as shown in fig. \ref{fig:Figure6} and fig. \ref{fig:Figure7}. When \(\beta\) is held constant at \(\beta=1\), the parameter \(\alpha\) must lie within: \(-2\times10^{80} \leq \alpha \leq -2 \times 10^{77}\), in a similar fashion, by  fixing \(\alpha\) at \(\alpha=-5\times10^{96}\), restricts  \(\beta\)  to the interval: \(-4.2\times10^{22} \leq \beta \leq -7.8\times10^{21}\). Under these constraints, the theoretically calculated value of BnER \((\frac{\eta}{s})\) reside within the range: \(7 \times 10^{-11} \leq \frac{\eta}{s} \leq 2\times 10^{-10}\), aligning remarkably well with the observed value of the matter imbalance existing in our universe. Fig. \ref{fig:Figure8} and fig. \ref{fig:Figure9} shows the variation of \(\frac{\eta}{s}\) as a function \(n\) for different values of \(\alpha\) and \(\beta\), respectively.

    \end{itemize}
    
    \item \textbf{Model II}: For Model II, we have taken, \(f(R)\) of the form \(f(R)=\alpha R^{\gamma} +\beta R^{\delta}\). We investigated the specified conditions: (\(\gamma\),\(\delta\)) replaced by (2,1) and (4,1), as two independent cases of study. For both the cases, we have plotted \(\frac{\eta}{s}\) as a function of \(\beta\) and \(n\).

    \begin{itemize}
    
    \item \textbf{Case I}: In this case, we have considered the range of \(n\) as: \(0.55 \leq n \leq 0.85\). By fixing \(\alpha=1\), \(\beta\) is constrained to lie within: \(9 \times 10^{44} \leq \beta \leq 1.5 \times 10^{46}\), for which the theoretically calculated asymmetry ratio lies in the range \(7 \times 10^{-11} \leq \frac{\eta}{s} \leq 2\times 10^{-10}\), which is highly in agreement to the observed value. Fig. \ref{fig:Figure11} and fig. \ref{fig:Figure12} shows the variation of \(\frac{\eta}{s}\) as a function \(\beta\) and \(n\) for different values of \(n\) and \(\beta\), respectively. 

    \item \textbf{Case II}: Here, we have taken \(n\) to lie within: \(0.55 \leq n \leq 0.85\). By fixing the value of \(\alpha\) at \(\alpha=10^{-25}\), we can have a viable range of values for \(\beta\) i.e. \(9 \times 10^{44} \leq \beta \leq 1.5 \times 10^{46}\), for which the theoretically calculated value of \(\frac{\eta}{s}\) is constrained to the interval: \(7 \times 10^{-11} \leq \frac{\eta}{s} \leq 2\times 10^{-10} \), that show great agreement with the astronomically obtained results. Fig. \ref{fig:Figure14} and fig. \ref{fig:Figure15} show the variation of \(\frac{\eta}{s}\) as a function \(\beta\) and \(n\) for different values of \(n\) and \(\beta\), respectively.

    \end{itemize}
    
    \item \textbf{Model III}: For Model-III, we have taken, \(f(R)\) of the form \(f(R)=\frac{\alpha R^{2m}-\beta R^m}{1+\gamma R^m}\). We have considered the case \(m=1\). We have plotted BnER, \(\frac{\eta}{s}\), as a function of \(\alpha\), \(\beta\) and \(n\).

    We have shown the variation of \(\frac{\eta}{s}\) as a function \(\alpha\) in fig. \ref{fig:Figure17} where \(n\) lies in within: \(0.55 \leq n \leq 0.85\). By fixing \(\beta=1\) and \(\gamma=1\times10^{-8}\), the matter excess ratio can be produced for the values of \(\alpha\) lying in the range: \(9\times10^{36} \leq \alpha \leq 1.5 \times 10^{38}\). In fig. \ref{fig:Figure18} we have taken range for \(n\) as: \(0.88 \leq n \leq 1.02\). By fixing, \(\alpha=1\) and \(\gamma=1\times10^{-48}\), the values of \(\beta\) lies within: \(9\times10^{45}\leq\beta\leq7\times10^{46}\), for which the calculated value of \(\frac{\eta}{s}\) (i.e., \(7.5\times10^{-11}\leq\frac{\eta}{s}\leq 3\times10^{-10} \)) matches to that of observed value with high accuracy. Fig. \ref{fig:Figure19} shows the variation of \(\frac{\eta}{s}\) as a function \(n\) for different values of \(\beta\).

\end{enumerate}

Fig. \ref{fig:Figure5} and Fig. \ref{fig:Figure10} represent the regions in the parameter space of \(\alpha\) and \(\beta\) for the two cases \(m=1\) and \(m=2\) of Model I, where \(8\times10^{-11}\leq \frac{\eta}{s} \leq 2\times10^{-10}\). Fig. \ref{fig:Figure13} and fig. \ref{fig:Figure16} displays the corresponding parameter space regions for the two cases of Model II, while Fig. \ref{fig:Figure20} illustrates the results for Model III under the same constraints. The shaded region of the parameter space shows the values of \(\alpha\) and \(\beta\) for which the BnER lies within the observational constraint range of \(8\times10^{-11}\leq \frac{\eta}{s} \leq 2\times10^{-10}\). 

Finally, the cosmological feasibility of the $f(R)$ gravity models under consideration needs to be explored. The models of $f(R)$ we have chosen to investigate GB can unify early inflation, radiation and matter-dominated epochs and late-time acceleration \cite{nojiri2006modified,nojiri2008modified}. The transition between these epochs is smooth and consistent with observations, where the effective cosmological constant at late times emerges naturally from the gravitational action rather than as an imposed parameter\cite{ nojiri2008modified}. In conclusion, the $f(R)$ theories employed in this study are able to unite all the key cosmological epochs that ensure the dynamical evolution of $R$. This dynamical transformation inherently generates the essential non-equilibrium as well as \(\mathcal{CP}\)-violation criteria needed for GB at different transitory points.

\bibliographystyle{apsrev4-2}
\bibliography{references}

\end{document}